\documentclass[a4paper,twoside]{article}
\usepackage[T1]{fontenc}
\usepackage{amsmath,amssymb,amsthm}

\newcommand{\open}[1]{{\mathbb{#1}}}
\newcommand{\signe}[1]{(-1)^{#1}}
\newcommand{\perm}{{\mathrm{perm}}}
\newcommand{\ee}{{\mathrm{e}}}
\newcommand{\dd}{{\mathrm{d}}}
\newcommand{\grad}{{\mathrm{deg}}}
\newcommand{\kbf}{{\mathbf{k}}}
\newcommand{\rbf}{{\mathbf{r}}}
\newcommand{\qbf}{{\mathbf{q}}}
\newcommand{\xbf}{{\mathbf{x}}}
\newcommand{\Lcal}{{\cal L}}
\newcommand{\Fcal}{{\cal F}}
\newcommand{\Ccal}{{\cal C}}
\newcommand{\ZZ}{{\partial\rho^{-1}}}
\newcommand{\tildu}{{\tilde{u}}}
\newcommand{\tildv}{{\tilde{v}}}
\newcommand{\counit}{{\varepsilon}}

\renewcommand{\i}[1]{{}_{\scriptscriptstyle(#1)}}
\newcommand{\ii}[1]{{}_{{\scriptscriptstyle{(}}{\scriptstyle{#1}}
          {\scriptscriptstyle{)}} }}
\newcommand{\iu}[1]{{}_{\scriptscriptstyle(\underline #1)}}
\newcommand{\id}{\mathrm{id}}
\newcommand{\tens}{\otimes}
\newcommand{\cou}{\counit}
\newcommand{\cop}{\Delta}
\newcommand{\antip}{\gamma}
\newcommand{\Sym}{{\mathsf{Sym}}}
\newcommand{\pwedge}{{\scriptstyle\vee}}
\newcommand{\barpsi}{{\overline{\psi}}} 
\newcommand{\barphi}{{\bar{\varphi}}}
\newcommand{\C}{\mathbb{C}}
\newcommand{\Zi}{\mathbb{Z}}
\newcommand{\Hom}{\mathrm{Hom}}
\newcommand{\defeq}{:=}
\newcommand{\HNcal}{{\hat{H}}}
\newcommand{\HN}{{H}}
\newcommand{\ANcal}{{\hat{A}_N}}
\newcommand{\AN}{{A_N}}
\newcommand{\AOcal}{{\hat{A}_O}}
\newcommand{\AO}{{A_O}}
\newcommand{\AT}{{A_T}}
\newcommand{\VN}{\hat{V}}
\newcommand{\contra}[2]{\,{\buildrel \,\hbox{$
    \vrule height 2.7pt width 0.5pt depth 0pt
    \vrule height 2.7pt width #1pt depth -2pt
    \vrule height 2.7pt width 0.5pt depth 0pt $}
    \over {#2} }\,}
\newcommand{\kon}[1]{\contra{6}{#1}}

\newtheorem{prop}{Proposition}[section]
\newtheorem{lem}[prop]{Lemma}
\newtheorem{cor}[prop]{Corollary}

\begin{document}
\begin{titlepage}
\title{\textsf{Quantum field theory and\\ Hopf algebra cohomology}}
\author{Christian Brouder*,
Bertfried Fauser$\sharp$,\\ Alessandra Frabetti\dag,
and Robert Oeckl\ddag\\ \\
*\ Laboratoire de Min\'eralogie-Cristallographie, CNRS UMR 7590,\\
 Universit\'es Paris 6 et 7, IPGP, 4 place Jussieu,\\
  F-75252 Paris Cedex 05, France\\ \\
$\sharp$ Universit\"at Konstanz,
   Fachbereich Physik, Fach M678,\\
   D-78457 Konstanz, Germany\\ \\
\dag\ Institut Girard Desargues, CNRS UMR 5028, \\
   Universit\'e de Lyon 1,
  21 av. Claude Bernard,\\ F-69622 Villeurbanne, France\\ \\
\ddag\ Centre de Physique Th\'eorique,
   CNRS UPR 7061,\\ F-13288 Marseille Cedex 9, France}

\date{CPT-2003/P.4618\\ April 13, 2004}
\maketitle
\begin{abstract}
We exhibit a Hopf superalgebra structure of the algebra of field
operators of quantum field theory (QFT) with the normal product.
Based on this we construct
the operator product and the time-ordered product as a twist
deformation in the sense of Drinfeld.
Our approach yields formulas for (perturbative) products and
expectation values that
allow for a significant enhancement in computational efficiency as
compared to traditional methods.
Employing Hopf algebra cohomology sheds new light on the
structure of QFT and 
allows the extension to interacting (not necessarily perturbative)
QFT. We give a
reconstruction theorem for time-ordered products in the spirit of
Streater and Wightman and recover the distinction between free and
interacting theory from a property of the underlying cocycle. We also
demonstrate how non-trivial vacua are described in our approach
solving a problem in quantum chemistry.

\end{abstract}

\end{titlepage}

\tableofcontents

\section{Introduction}

The purpose of this article is to present a new approach to the
algebraic and combinatorial structures at the heart of quantum field
theory (QFT). This approach has merits on the practical as well as
on the conceptual side. On the practical side it allows for a major
computational enhancement based on an efficient description of the
combinatorics and on non-recursive closed formulae. On the conceptual
side it gives new insights into the algebraic structure of QFT. We
evidence this through applications to
non-perturbative QFT and non-trivial vacua.

The starting point of our approach is the identification of a Hopf
algebraic structure at the core of QFT. That is, the algebra of
field operators with the normal product is a \emph{Hopf
superalgebra}. This
means that besides the product there is a \emph{coproduct} that
describes, intuitively speaking, the different ways in which a product
of field operators might be partitioned into two sets. Indeed, it is
this coproduct that plays the key role in a closed description of
combinatorial structures and that allows for computationally efficient
algorithms. Another key structure of the Hopf superalgebra is the
\emph{counit}. This turns out to describe the standard vacuum
expectation value. Algebraically, this Hopf superalgebra is the graded
symmetric Hopf algebra (described in detail in
Appendix~\ref{sec:Example}, see also \cite[Appendix~2]{Eisenbud}).
The conceptual origin of this Hopf superalgebra is rather
simple. Identifying the normal ordered products with functionals on
field configurations, the coproduct is induced by the linear addition
of fields.

The second main step consists in identifying the standard canonical
quantization with a
twist in the sense of Drinfeld \cite{Drinfeld}. More precisely, the
operator product emerges as a twist deformation of the normal
product. As is common we deal here at first with the free QFT.
The twist is induced by a \emph{Laplace pairing} which in
turn is determined by a suitable propagator.
Furthermore, the
time-ordered product can be obtained similarly as a direct twist
deformation of the normal product. In this case, the Laplace pairing
is determined by the Feynman propagator. 
Since vacuum expectation values
of time-ordered products are the main ingredients of physical
scattering amplitudes this allows the use of our methods in actual
calculations of physical quantities.

It is one of the basic facts in quantum field theory that Wick's
theorem relates normal and time-ordered correlation functions 
\cite{Wick,Anderson,Stumpf}. It was only recently noticed by Fauser
that this  transformation can be advantageously described in Hopf
algebraic terms \cite{Fauser,Fauser98,FauserHabi}. This is indeed a
crucial ingredient of our construction, which allows us to prove that
the twists yield the desired products. In particular, we show that
the Hopf algebraic Wick transformation can be applied to yield the
operator product
from the normal product in a way analogous as for the time-ordered
product.

Other aspects of our approach described so far are also present in the
literature in different guises.
Oeckl used the (dual of the) present Hopf algebra structure to
generalize QFT \cite{Oeckl,OecklPhD} to quantum group symmetries and
the twist to describe QFT on noncommutative spaces \cite{Oe:nctwist}.
Borcherds defined vertex algebras with quantum group
methods \cite{Borcherds,Borcherds2}. In particular, he uses a
construction similar to the twist in a related context.
The research of the present
paper was initiated by Brouder \cite{BrouderQFA} and a
preliminary presentation of parts of this work was given in
\cite{BrouderOecklI}.

The closed formulae emerging in our framework facilitate highly
efficient computations of products and expectation values. This is
mainly due to the heavy use of the coproduct structure. Indeed, this is
well known in combinatorics where Hopf algebraic methods are
established for this purpose.
We know from computer algebra calculations that precisely the
techniques employed enhance performance and that in a well defined sense
no algorithm can come up with 
less terms, see \cite{AF1,AF2}.
While the twisted products described so far are the products of the
free QFT our
framework is naturally compatible with the usual perturbation theory
and thus applicable to it. This implies that the computational
advantages directly apply to perturbative QFT.

The third step consists in exploiting the Hopf algebra
cohomology theory due to Sweedler, which underlies the twisted product
\cite{Sweedler}. (In that
work twisted products of the type used here where also introduced for
the first time.) Besides affording conceptual insight this yields
immediate practical benefits. Among these is the realization of the
time-ordering prescription of QFT as an algebra isomorphism. This in
turn can be used on the computational side.

The cohomological point of view affords a further extension of our
framework. Significantly, twisted products can not only be defined
with Laplace pairings, but with 2-cocycles, of which Laplace pairings
are only a special case. Remarkably, it turns out that 2-cocycles that
are not Laplace pairings lead to (non-perturbatively described)
interacting QFTs. Moreover, any QFT (with
polynomial fields) can be obtained in this way. A QFT is free if and
only if the 2-cocycle is a Laplace pairing.

A further application of the cohomology that we develop is to
non-trivial vacua. We show that changing the choice of vacuum can also
be encoded through a twist. Indeed, it turns out that there is a
``duality'' or correspondence between the choice of vacuum and that of
product. We exemplify this result by solving a problem posed by
Kutzelnigg and Mukherjee \cite{Kutzelnigg} regarding ``adapted normal
products'' in quantum chemistry. While they were able to
give only examples for low orders, our framework yields closed
formulas for all orders. Our method is capable of describing
condensates too, as we know from \cite{FauserVacua}.

Although we do not develop this point of view in the present paper, a
twist in the sense used here is automatically an (equivariant)
deformation quantization. Indeed, this was one of the original
motivations for Drinfeld to introduce this concept
\cite{Dri:constybe}.
This means that our approach is thus inherently connected to the
deformation quantization approach to QFT. This approach starts also
with the normal ordered product and views the other products
as deformations of
it. See the recent paper by Hirshfeld and Henselder
\cite{HiHe:defquantqft}.

The paper is roughly divided into three parts. The first, consisting of
Section~\ref{sec:2} starts by introducing a few essential mathematical
concepts. Then, the Hopf algebra structure of the normal ordered field
operators is developed. Next, the operator and time-ordered
products are constructed as twisted products induced by Laplace
pairings. We finish the section with closed formulas for Wick
expansions, the various products and expectation values showing the
practical efficiency of the framework. This part of the paper is
intended for a broad audience and should be readable without prior
familiarity with Hopf algebras.

The second part of the paper consists of Section~\ref{sec:3}. Here we
go deeper into the underlying mathematics, starting with a brief
review of Hopf algebra cohomology and Drinfeld twisting. Then we turn
to the implications for QFT. In particular, we describe the
cohomological understanding of the operator and time-ordered product
as twisted products. Among other practical consequences we derive the
time-ordering operation as an algebra isomorphism.

The third part of the paper presents further results emerging from the
cohomological insights. It consists of Sections~\ref{sec:inter} and
\ref{sec:vacua}. In Section~\ref{sec:inter} interactions are
treated. Firstly, we show that our framework is compatible with
perturbation theory and thus allows the application of our methods
there. Secondly, we consider general (and not necessarily
perturbative) quantum field theory and show that our framework
naturally extends to it. In particular, we present the reconstruction
theorem that allows to describe any (linear and polynomial) QFT
through a 2-cocycle. In Section~\ref{sec:vacua} we show how choosing
non-trivial vacua can be naturally expressed in our framework. Moreover,
we demonstrate the efficient solution of a problem arising in quantum
chemistry with our approach.

After conclusions and outlook the paper ends with two
Appendices. Appendix~\ref{sec:superalg} gives some elementary
definitions on Hopf superalgebras and in particular the graded
symmetric Hopf superalgebra that plays the crucial role in this
paper. The terminology of Hopf $*$-superalgebras is not unique and
even in general incompatible between different sources. So a
further value of this appendix is that it collects in a coherent and
compatible way notions scattered in the literature.
Appendix~\ref{cohomsect} consists firstly of a short description
of the cohomology groups in the bosonic case and secondly of the more
technical proofs of Propositions and Lemmas appearing in the main
text.

We refer readers who wish to know more about Hopf algebras and quantum
group theory to \cite{Kassel} and \cite{Majid}. The latter reference is
particularly suitable for the cohomology theory and the twist
construction.

\section{Free quantum field theory}
\label{sec:2}

\subsection{Mathematical basis}

We start in this section by introducing a few mathematical concepts
that will be required throughout the article. These are, apart from
Hopf (super)algebras, the Laplace pairing and the twisted product. It
should be possible even for the reader without previous experience
with Hopf algebras to follow the main steps of
Section~\ref{sec:2}. Indeed, a
first reading should be possible without paying too much attention to
the details of definitions.

\subsubsection{Hopf $*$-superalgebra}

Recall that a \emph{Hopf algebra} $H$, besides being an associative algebra
with a unit, has a coassoiative \emph{coproduct} $\cop:H\to H\tens H$, a
\emph{counit} 
$\cou:H\to \C$ and an \emph{antipode} $\antip: H\to H$, satisfying
compatibility conditions. By definition the coproduct of an element
$a$ of $H$ can always be written as a (non-unique) linear combination
of the form $\cop a =\sum_i a'_i\tens a''_i$. In order to
avoid the proliferation of indices it is customary to use \emph{Sweedler's
notation} for this, i.e.\ we write $\cop a =\sum a\i1\tens a\i2$.
Due to the coassociativity Sweedler's notation extends
unambiguously to multiple coproducts as follows: $\sum (a\i1)\i1\tens
(a\i1)\i2 \tens a\i2 = \sum a\i1\tens
(a\i2)\i1 \tens (a\i2)\i2 = \sum a\i1\tens a\i2\tens a\i3$ etc.

A
superalgebra $A$ is a $\Zi_2$-graded
algebra so that $|a b|=|a| + |b|$ modulo 2 where $|a|$ denotes the
parity of the
element $a$, $|a|=0$ if it is even (bosonic) and $|a|=1$ if it is odd
(fermionic). A \emph{Hopf superalgebra} is a superalgebra with unit
and $\Zi_2$-graded coproduct, counit and antipode. A
\emph{$*$-(super)algebra} is a
(super)algebra $A$ with an antilinear map $*:A\to A$ such that $(a
b)^*=b^* a^*$. A Hopf $*$-superalgebra is a Hopf superalgebra and
$*$-superalgebra where the $*$-structure is compatible with coproduct,
counit and antipode.

We refer the reader who is not familiar with these notions or wishes
to recall their details to the Appendix~\ref{sec:superalg}, where the
complete definitions are given on an elementary level.

\subsubsection{Laplace pairing}
\label{sec:deflap}

We introduce now the concept
of Laplace pairing which is relevant to interpret
Wick's theorem in terms of Hopf superalgebras.

Let $H$ be a Hopf superalgebra that we require to be graded
cocommutative. That is, the coproduct satisfies $a\i1\tens a\i2=
(-1)^{|a\i1| |a\i2|} a\i2\tens a\i1$.
A \emph{pairing} on $H$ is a linear map
$(\cdot|\cdot) :  H \otimes H \rightarrow \mathbb{C}$.
It is called even if $(a|b)=0$ when the parities
of $a$ and $b$ are different.
A \emph{Laplace pairing}\footnote{
This name was given by Rota some time ago \cite{Doubilet, Grosshans},
because equations (\ref{eq:pair1}) and (\ref{eq:pair2}) are an elegant
way of writing the Laplace identities for determinants. Equation
(\ref{eq:pair1}) is called expansion by rows and equation
(\ref{eq:pair2}) expansion by columns. They express the
determinant in terms of minors (see Ref.~\cite{Vein} p.26 and
Ref.~\cite{Muir} p.93) and were derived by Laplace in 1772
\cite{Laplace}.}
on $H$ is an even pairing on $H$ such that the product
and the coproduct are dual in the sense that
\begin{eqnarray}
(a a'|b) &=& \sum \signe{|a'|\,|b\i1|} (a|b\i1 )(a'|b\i2 ),
\label{eq:pair1} \\
(a|bb') &=& \sum \signe{|a\i2 |\,|b|} (a\i1|b)(a\i2 |b'), \label{eq:pair2}
\end{eqnarray}
and the unit and counit are dual as follows,
\begin{eqnarray}
(1|a)=(a|1) &=& \cou(a).
\label{eq:pair3}
\end{eqnarray}

\subsubsection{Twisted product}

A Laplace pairing on $H$ can be used to deform 
the product of $H$. Sweedler \cite{Sweedler}
defined the twisted product on $H$, which we denote by $\circ$,
as\footnote{Sweedler considered only the bosonic case. He called the
  product a crossed product and his definition was somewhat more
  general (in a different direction).}
\begin{eqnarray}
a\circ b &=&
\sum \signe{|a\i2|\,|b\i1|}\,
  (a\i1|b\i1) a\i2 b\i2
\label{defcircle}
\end{eqnarray}
The twisted product $\circ$ is associative, and $1$ is also the unit
for $\circ$.
As we shall see later, this twisted product yields an
elegant way to write Wick's theorem. From the mathematical
point of view, this arbitrary seeming definition 
can be understood as a special case of the more fundamental
Drinfeld twist discussed in Section~\ref{sec:twist}.

Notice also that the twisted product is not compatible
with the coproduct: in general
$\Delta (a\circ b)\not= \sum (-1)^{|b\i1||a\i2|}
(a\i1\circ b\i1) \otimes (a\i2\circ b\i2)$.

\subsection{The Hopf superalgebra of creation and annihilation operators}

In this section, we define the Hopf superalgebra of creation
and annihilation operators. The fermion and boson creation and annihilation
operators will be treated in a unified way.
Normal products of creation and annihilation operators 
form a well-known graded commutative superalgebra,
called the symmetric superalgebra (see Appendix~\ref{sec:superalg}).
In this section, we define a coproduct and a counit
which are compatible with the normal product,
and we equip the normal products of operators 
with the structure of a Hopf superalgebra.
The Hopf superalgebra of creation and annihilation operators
will be used to define the Hopf superalgebra of quantum fields.

We first denote by $\varphi(x;s)$ the solutions of the
classical field equations (e.g. the classical Klein-Gordon, Dirac
or Maxwell equations). The solution $\varphi(x;s)$
is a function of the spacetime variable $x$ 
and $s$ indexes the solutions of the classical
field equations.
In the vacuum, $s$ is the 3-momentum $\kbf$ for scalar
fields. For Dirac fields there are positive energy solutions
$\varphi_>(x,s)$ with $s=(\kbf,\alpha)$ and
$\alpha=1,2$ for up and down spin states
and negative energy solutions 
$\varphi_<(x,s)$ with $s=(\kbf,\alpha)$ and
$\alpha=1,2$ for up and down spin states.
For the classical field equations in an external potential,
$s$ is a discrete index for the bound state
and a continuous index for the scattering states.
For $x=(0,\rbf)$, the set of functions $\varphi(x;s)$ is
assumed to form a suitable space of functions of $\rbf$. 
The functions $\varphi(x;s)$ will be used to define
quantum fields.

\subsubsection{The superalgebra structure}

The creation and annihilation operators are
denoted by $a^\dagger(s)$ and $a(s)$, respectively.
They create and annihilate a particle in the state $s$.
They are operators acting on a Fock space $\Fcal$ and their
precise definition is given in ref.~\cite{ReedSimonII} p.218.
The normalized state of $\Fcal$ corresponding to no particles is 
called the vacuum and denoted by $|0\rangle$.
The parity of these operators is $|a(s)|=|a^\dagger(s)|=1$
for a fermion field and $|a(s)|=|a^\dagger(s)|=0$
for a boson field.
The operator product of two operators $u$ and $v$
is written $uv$. The (anti)commutation
relations among creation operators and among annihilation operators of
bosons (fermions) \cite{Itzykson} can be summarized as
\begin{eqnarray}
a(s)a(s') &=& (-1)^{|a(s)||a(s')|} a(s')a(s),\nonumber\\
a^\dagger(s)a^\dagger(s') &=& (-1)^{|a^\dagger(s)||a^\dagger(s')|} 
a^\dagger(s')a^\dagger(s).
\label{gradedcom}
\end{eqnarray}
These equations mean that two annihilation operators or
two creation operators commute
for bosons, anticommute for fermions and
commute for a boson and a fermion.

The superalgebra $\ANcal$ of normal products is generated as a vector space
by products of creation operators on the left of 
products of annihilation operators.
For example $u= a^\dagger(s_1)\dots a^\dagger(s_m)
a(s_{m+1})\dots a(s_{m+n})$ is an element of $\ANcal$.
The parity of $u$ is
$|u|=|a^\dagger(s_1)|+\dots+|a^\dagger(s_m)|+
|a(s_{m+1})|+\dots+|a(s_{m+n})|$ modulo 2.
The element given by $m=0$ and $n=0$ in this example is
the unit operator denoted by $1$.
The product in $\ANcal$ is the normal product.
In quantum field theory, the normal product of
two elements $u$ and $v$ is denoted by ${:}uv{:}$
but this notation becomes cumbersome when we manipulate
various products of several fields, so we prefer
to denote the normal product by $u\pwedge v$, which is
the standard mathematical notation for a graded-commutative product.
The normal product is defined by 
$a^\dagger(s) \pwedge a^\dagger(s')=a^\dagger(s) a^\dagger(s')$,
$a^\dagger(s) \pwedge a(s')=a^\dagger(s) a(s')$,
$a(s) \pwedge a^\dagger(s')=(-1)^{|a(s)||a^\dagger(s')|} a^\dagger(s') a(s)$,
$a(s) \pwedge a(s')=a(s) a(s')$ and extended to $\ANcal$ by associativity
and linearity.
From the definition of the normal product and the
relations (\ref{gradedcom}), we see that if $u$ and $v$ are in $\ANcal$, 
$u\pwedge v=(-1)^{|u||v|}v\pwedge u$. That is, the normal product
is graded commutative.
Hence, the superalgebra $\ANcal$ is a graded-commutative associative
superalgebra
with unit $1$. These results can be summarized in the following
proposition
\begin{prop}
\label{isomo}
If $\VN$ is the vector space generated by $a(s)$ and $a^\dagger(s)$
(for all $s$),
the superalgebra $\ANcal$ of normal products has the structure of the
symmetric superalgebra $\Sym(\VN)$.
\end{prop}
The symmetric superalgebra $\Sym(\VN)$ is described in 
Appendix~\ref{sec:superalg}.
If the theory contains bosons and fermions
the vector space $\VN$ generated by $a(s)$ and $a^\dagger(s)$
for all $s$ can be written as $\VN=\VN_0\oplus \VN_1$, 
where $\VN_0$ is generated by
the boson operators and $\VN_1$ by the fermion operators.

From Appendix~\ref{sec:superalg} we know that
$\Sym(\VN)$ has the structure of a Hopf superalgebra.
Thus, the superalgebra of creation and annihilation operators
has a Hopf superalgebra structure that will be discussed
in the next section. For later convenience, we distinguish
the superalgebra $\ANcal$ of creation and annihilation operators
without the full Hopf structure, and the Hopf superalgebra
of creation and annihilation operators, that we denote
by $\HNcal$.

\subsubsection{The Hopf $*$-superalgebra structure}

Starting from the Hopf superalgebra $\Sym(\VN)$, we see
that the coproduct of the Hopf superalgebra $\HNcal$ of creation
and annihilation operators is 
defined by $\Delta 1=1\otimes 1$,
$\Delta a(s)=a(s)\otimes 1+ 1\otimes a(s)$,
$\Delta a^\dagger(s)=a^\dagger(s)\otimes 1+ 1\otimes a^\dagger(s)$
on $\VN$ and extended to $\HNcal$ by
$\Delta (u\pwedge v)=\sum (-1)^{|u\i2||v\i1|}
(u\i1\pwedge v\i1)\otimes (u\i2\pwedge v\i2)$.
For example, if $a=a(s)$, $b=a(s')$ and $c=a(s'')$,
\begin{eqnarray*}
\Delta (a\pwedge b)&=&
(a\pwedge b)\otimes 1 + a \otimes  b + (-1)^{|a||b|}
 b \otimes a + 1 \otimes (a \pwedge b),\\
\Delta (a\pwedge b\pwedge c) &=&
 1\otimes a\pwedge b\pwedge c
 +a \otimes b\pwedge c
 +(-1)^{|a||b|} b \otimes a\pwedge c
\\&&
 +(-1)^{|a||c|+|b||c|}c \otimes a\pwedge b
 +a\pwedge b \otimes c
 +(-1)^{|b||c|}a\pwedge c \otimes b
\\&&
 +(-1)^{|a||b|+|a||c|}b\pwedge c \otimes a
+a\pwedge b\pwedge c\otimes 1.
\end{eqnarray*}
In general
\begin{eqnarray*}
\Delta (u^1\pwedge\dots \pwedge u^n) &=&
\sum (-1)^F\,
u\i1^1\pwedge\dots\pwedge u\i1^n
\otimes
u\i2^1\pwedge\dots\pwedge u\i2^n
\end{eqnarray*}
for any $u^1,\dots,u^n\in \HNcal$
and with
$F=\sum_{k=2}^n \sum_{l=1}^{k-1} |u\i1^k||u\i2^l|$.

In particular, if $a_1,\dots,a_n$ are
creation or annihilation operators,
the coproduct of 
$a_1\pwedge\dots\pwedge a_n$ is
given by equation (\ref{Deltav1vn}) of
Appendix~\ref{sec:Example}.

The counit of $\HNcal$ is defined by
$\counit(1)=1$, $\counit(a(s))=0$
and $\counit(a^\dagger(s))=0$ and extended to
$\HNcal$ by
$\counit(u\pwedge v)=\counit(u)\counit(v)$ for
any $u$ and $v$ in $\HNcal$.
Therefore $\counit(u)=0$ if
$u=a^\dagger(s_1)\dots a^\dagger(s_m)
a(s_{m+1})\dots a(s_{m+n})$ for
$m>0$ or $n>0$. 
The relation between Hopf algebra and quantum field
concepts is strengthened  by the following
\begin{prop}
\label{prop:vev}
For any normal product, i.e. any element
$u\in\HNcal$, the counit is equal to the
vacuum expectation value:
$\counit(u)=\langle 0|u|0\rangle$.
\end{prop}
To show this, we evaluate $\epsilon(u)$
and $\langle 0|u|0\rangle$ for all elements
of a basis of $\HNcal$.
The proposition is true for the unit because
$\counit(1)=1=\langle 0|1|0\rangle$.
Take now a basis element of $\HNcal$
$u=a^\dagger(s_1)\dots a^\dagger(s_m)
a(s_{m+1})\dots a(s_{m+n})$ for
$m>0$ or $n>0$. Then
$\counit(u)=0=\langle 0|u|0\rangle$.
The result follows for all elements of
$\HNcal$ by linearity of the counit and
of the vacuum expectation value.
This relation between the counit and the expectation
value over the vacuum was already pointed out in \cite{Fauser}.

To complete the description of the
Hopf superalgebra $\HNcal$, we define its antipode by
$\antip\big(a^\#(s_1)\pwedge\dots\pwedge a^\#(s_n)\big)
=
(-1)^n a^\#(s_1)\pwedge\dots\pwedge a^\#(s_n)$,
where $a^\#(s_i)$ stands for $a^\dagger(s_i)$ or $a(s_i)$.
Moreover, $\HNcal$ has a $*$-structure
generated by $a(s)^*=a^\dagger(s)$.

\subsection{The Hopf superalgebra of field operators}
\label{sec:salgfields}

The operators used in the superalgebra $\ANcal$ of normal products
are independent of space and time, they are indexed by
the solutions of the classical equation.
Now we introduce space- and time-dependent field operators
for Dirac and scalar fields.
An excellent description of field operators can be
found in \cite{ReedSimonII}.

\subsubsection{The field operators}
To define the Dirac field operator, we need to split the set of
solutions of the Dirac equation into two groups, the positive energy 
states $\varphi_>(x;s)$ and the negative energy states
$\varphi_<(x,s)$.
The solutions with positive energy are
$\varphi_>(x;n)$ with energy $E_n<m$ (where $m$ is the
electron mass) for bound states
and $\varphi_>(x;\kbf,\alpha)$ with energy
$\omega_k=\sqrt{\kbf\cdot\kbf+m^2}$
for continuum states. The solutions
with negative energy are assumed to be always
continuum states  $\varphi_<(x;\kbf,\alpha)$
with energy $-\omega_k$.

The Dirac field operator is defined by \cite{Itzykson}
\begin{eqnarray*}
\psi(x) &=& \sum_n \varphi_>(x;n) b_n
   + \int \dd\mu(\kbf)\sum_{\alpha=1}^2  
          \varphi_>(x;\kbf,\alpha) b_\alpha(\kbf)
   +      \varphi_<(x;\kbf,\alpha) d^\dagger_\alpha(\kbf).
\end{eqnarray*}
In this expression, $\dd\mu(\kbf)= m/(8\pi^3 \omega_k)\dd\kbf$.
Its Dirac adjoint is
\begin{eqnarray*}
\barpsi(x) &=& \sum_n \barphi_>(x;n) b^\dagger_n
   + \int \dd\mu(\kbf)\sum_{\alpha=1}^2  
          \barphi_>(x;\kbf,\alpha) b^\dagger_\alpha(\kbf)
   +      \barphi_<(x;\kbf,\alpha) d_\alpha(\kbf).
\end{eqnarray*}
with $\barphi=\varphi^\dagger \gamma^0$ \cite{Itzykson},
where $\varphi^\dagger$ is the adjoint of the
spinor $\varphi$. The creation and annihilation operators
are $b^\dagger_n$ and $b_n$ for bound states,
$b^\dagger_\alpha(\kbf)$ and $b_\alpha(\kbf)$
for positive energy scattering states and
$d^\dagger_\alpha(\kbf)$ and $d_\alpha(\kbf)$
for negative energy scattering states.

For a neutral scalar field $\phi(x)$ in the vacuum, the construction
is simpler
\begin{eqnarray*}
\phi(x) &=& \int \big(\varphi(x;\kbf) a(\kbf)
+\varphi^\dagger(x;\kbf) a^\dagger(\kbf) \big)\dd \mu(\kbf).
\end{eqnarray*}

\subsubsection{The Hopf superalgebra of fields}
We define $V$ as the vector space generated by
the free fields
(e.g. $\psi(x)$, $\barpsi(x)$ and $A^\mu(x)$ for all $x$
in quantum electrodynamics).
Then the Hopf superalgebra $\HNcal=\Sym(\VN)$ extends to a Hopf 
superalgebra structure on $\HN=\Sym(V)$.
The normal product of creation and annihilation operators
extends to a normal product of fields, also denoted by
$\pwedge$. For example 
\begin{eqnarray*}
\phi(x)\pwedge \phi(y) &=& 
\int  \dd \mu(\kbf)\dd \mu(\qbf)
 \big(\varphi^\dagger(x;\kbf) \varphi(y;\qbf) 
        a^\dagger(\kbf)\pwedge a(\qbf) 
\\ &&
+ \varphi^\dagger(x;\kbf) \varphi^\dagger(y;\qbf) 
        a^\dagger(\kbf) \pwedge  a^\dagger(\qbf)
+
 \varphi(x;\kbf) \varphi(y;\qbf) 
        a(\kbf) \pwedge a(\qbf)
\\ &&
+ \varphi(x;\kbf) \varphi^\dagger(y;\qbf) 
        a(\kbf) \pwedge a^\dagger(\qbf) \big).
\end{eqnarray*}

The coproduct is extended from the coproduct of $\HNcal$.
This extension uses the fact that the transformation from
$\VN$ to $V$ is linear. For example
\begin{eqnarray*}
\Delta A_\mu(x) &=& A_\mu(x) \otimes 1 + 1\otimes A_\mu(x),\\
\Delta \psi(x) &=& \psi(x) \otimes 1 + 1\otimes \psi(x),\\
\Delta \big(\barpsi(x)\pwedge\psi(y)\big) &=& 
  \barpsi(x)\pwedge\psi(y) \otimes 1 +
   1\otimes \barpsi(x)\pwedge\psi(y)
     + \barpsi(x)\otimes \psi(y)
\\&&
     - \psi(y)\otimes \barpsi(x).
\end{eqnarray*}
The counit of $\HN$ is extended from the counit of $\HNcal$.
Thus,
$\counit(\phi(x))=\counit(\psi(x))=\counit(\barpsi(x))=0$.
The antipode is defined by $\antip(u)=(-1)^n u$
if $u$ is the normal product of $n$ elements of $V$,
and extended to $\HN$ by linearity.
$\HN$ is a graded commutative and graded cocommutative
Hopf superalgebra.

The $*$-structure of quantum field operators is
deduced from the $*$-structure on creation and
annihilation operators $a(s)^*=a^\dagger(s)$.
It gives $\phi(x)^*=\phi(x)$ for a neutral
scalar field and
$\psi(x)^*=\psi^\dagger(x)=\barpsi(x)\gamma^0$
and $\barpsi(x)^*=\gamma^0\psi(x)$
for Dirac fields.
The $*$-structure is related to the
charge conjugation operator $\Ccal$
of Dirac fields by
$\Ccal \psi(x) \Ccal^\dagger = i \gamma^2 \psi(x)^*$
(see \cite{Itzykson}).

\subsection{Operator and time-ordered product of field operators}

\label{sec:opto}

In this section we show that, by properly choosing the Laplace pairing,
we can twist the normal product into the operator product or the
time-ordered product.
The twisted product on $\HNcal$ or $H$
is given by equation (\ref{defcircle}).
From the coproduct of these Hopf superalgebras
and the definition (\ref{defcircle}) of the twisted
product, we obtain the following simple examples,
valid for $a,b$ and $c$ in $\VN$ (or $V$).
\begin{eqnarray}
a\circ b &=& a\pwedge b + (a|b), \label{acircb} \\
(a\pwedge b)\circ c &=& a\pwedge b \pwedge c +(-1)^{|b||c|} (a|c)b + 
   (-1)^{|a||b|+|a||c|}(b|c) a, \nonumber\\
a\circ ( b \pwedge c) &=& a\pwedge b \pwedge c +(-1)^{|b||c|} (a|c)b +
(a|b) c,
\nonumber\\
a\circ b \circ c &=& a\pwedge b \pwedge c + (a|b) c +(-1)^{|b||c|} (a|c)b
+(-1)^{|a||b|+|a||c|}(b|c) a.
\nonumber
\end{eqnarray}
Now we are going to specify the Laplace pairings relevant to the
algebra of normal products $\HNcal$ and the algebra of fields $H$.

\subsubsection{The algebra of operator products $\AOcal$} 
We call \emph{Wightman pairing} the Laplace pairing defined as follows.
For a scalar field in the vacuum the Wightman pairing is 
\begin{eqnarray}
(a(\kbf)|a^\dagger(\qbf))_+ &=& 
\frac{\delta(\kbf-\qbf)}{\rho(\kbf)},
\label{pairingbakaq}
\end{eqnarray}
where $\rho(\kbf)=(2\pi)^{-3} m/\sqrt{\kbf\cdot\kbf+m^2}$,
all the other pairings being zero.
For Dirac fields we have
\begin{eqnarray*}
(b_n|b^\dagger_p)_+ &=& \delta_{np},\\
(b_\alpha(\kbf)|b^\dagger_\beta(\qbf))_+ &=& 
(d_\alpha(\kbf)|d^\dagger_\beta(\qbf))_+ = 
\delta_{\alpha\beta} \frac{\delta(\kbf-\qbf)}{\rho(\kbf)},
\end{eqnarray*}
all the other pairings being zero.
This Wightman pairing twists the normal product into
the operator product and the algebra $\ANcal$ becomes 
the algebra $\AOcal$.
But we need first to prove the following proposition
\begin{prop}
\label{proofucircv}
The twisted product defined by the Wightman pairing
is equal to the operator product: for
any elements $u$ and $v$ of $\ANcal$, 
$u\circ v=uv$.
\end{prop}

For the case of a scalar field, we are going to 
show that the operator product of two elements $u$ and $v$
of $\ANcal$ is equal to the twisted product of these elements
with the Wightman pairing (\ref{pairingbakaq}).
The proof for Dirac fields is analogous.

For two operators $a$ and $b$, where $a=a(s)$ or $a=a^\dagger(s)$
 and $b=a(s)$ or $b=a^\dagger(s')$, equation (\ref{acircb}) tells
us that $a\circ b= a\pwedge b+ (a|b)_+$. On the other hand
we know from Wick's theorem \cite{Itzykson}
that the operator product satisfies
$a b= a\pwedge b + \langle 0|a b|0\rangle$ (recall
that $a\pwedge b ={:}ab{:}$). The Wightman pairing
was defined precisely so that
$(a|b)_+=\langle 0|a b|0\rangle$, thus 
$a\circ b=ab$.
This equality is valid for any elements $a$ and $b$ in $\VN$,
the vector space generated by $a(s)$ and $a^\dagger(s)$.

We must now prove that $u v=u\circ v$ for any $u$ and $v$ in $\ANcal$.
This is done by using Wick's theorem \cite{Wick}.
Wick's theorem is very well known, so we recall it only briefly.
It states that the operator
product of some elements of $\VN$
is equal to the sum over all possible pairs of
contractions (see e.g. \cite{Fetter} p.209,
\cite{WeinbergQFT} p.261, \cite{Ticciati} p.85).
A contraction\footnote{Contractions were first used
by Houriet and Kind \cite{Houriet}.} 
$\kon{a b}$ is the difference between the operator
product and the normal product. 
$\kon{ab}= ab-a\pwedge b$,
so that $\kon{ab}= \langle 0|ab|0\rangle =(a|b)_+$.

If $u=a_1\pwedge\dots \pwedge a_n$,
Wick's theorem for bosons
is proved recursively from the following identity
\cite{Wick}:
\begin{eqnarray}
u b &=& u \pwedge b
+ \sum_{j=1}^n (a_j|b)_+\,a_1\pwedge\dots\pwedge a_{j-1}\pwedge
a_{j+1}\pwedge\dots\pwedge a_n.
\label{Wickid}
\end{eqnarray}
Thus, to show that $u\circ b=ub$ we must recover
equation (\ref{Wickid}) from our definition.
In other words, we must prove
\begin{eqnarray}
u\circ b &=& u \pwedge b
+ \sum_{j=1}^n (a_j|b)_+\,a_1\pwedge\dots\pwedge a_{j-1}\pwedge
a_{j+1}\pwedge\dots\pwedge a_n.
\label{ucircb}
\end{eqnarray}
To show this, we make a recursive proof with respect to
the degree of $u$. We recall that an element has
degree $k$ if it can be written as the normal product of
$k$ creation or annihilation operators
(see section \ref{sec:Hopfsuperalgebras}).
We use the definition
(\ref{defcircle}) of the twisted product and the fact that
$\Delta b = b\otimes 1 + 1 \otimes b$ to find
\begin{eqnarray}
u\circ b &=& u \pwedge b + \sum (u\i1|b)_+ u\i2.
\label{ucircb2}
\end{eqnarray}
The Wightman pairing $(u\i1|b)_+$ is zero if 
 $u\i1$ is not of degree 1.
According to equation (\ref{Deltav1vn}) for $\Delta u$, this
happens only for the $(1,n-1)$-shuffles.
By definition, a $(1,n-1)$-shuffle is a permutation
$\sigma$ of $(1,\dots,n)$ such that
$\sigma(2)<\dots <\sigma(n)$, and the corresponding
terms in the coproduct of $\Delta u$
are
\begin{eqnarray*}
\sum_{j=1}^n a_j \otimes a_1\pwedge\dots\pwedge a_{j-1}\pwedge
a_{j+1}\pwedge\dots\pwedge a_n.
\end{eqnarray*}
Thus, we recover (\ref{ucircb}) from the Laplace identity
(\ref{ucircb2}) and the twisted product
of $u$ and $b$ is the operator product of $u$ and $b$.
By linearity of the twisted and operator products, this
can be extended to any element of $\ANcal$ and we have
$u\circ b=ub$ for any $u$ in $\ANcal$ and any $b\in \VN$. 
A similar argument leads to
\begin{eqnarray}
a\circ u &=& au, 
\label{acircu}
\end{eqnarray}
for any $a$ in $\VN$ and any $u$ in $\ANcal$.
Now we proceed by induction.
Assume that $u\circ v=uv$ for any $v\in\ANcal$ and for  $u$ 
of degree $k\le n$. We take now an element $u$ of degree $n$
and we calculate $(a\pwedge u)\circ v$ where $a$ is in $\VN$.
We use $a\pwedge u=a\circ u-\sum (a|u\i1)_+ u\i2$
and we write
\begin{eqnarray*}
(a\pwedge u)\circ v &=& (a\circ u)\circ v - \sum (a|u\i1)_+ u\i2\circ v\\
&=& a\circ (u\circ v) - \sum (a|u\i1)_+ u\i2\circ v\\
&=& a\circ (u v) - \sum (a|u\i1)_+ u\i2 v
= a u v - \sum (a|u\i1)_+ u\i2 v,
\end{eqnarray*}
by associativity of the twisted product and because of the recursion
hypothesis and equation (\ref{acircu}).
By associativity of the operator product this can be rewritten
\begin{eqnarray*}
(a\pwedge u)\circ v 
&=& (a u  - \sum (a|u\i1)_+ u\i2)v
= (a\circ  u  - \sum (a|u\i1)_+ u\i2)v
\\&=& (a\pwedge u)v.
\end{eqnarray*}
Therefore $u\circ v=uv$ if $u$ is 
of degree $n+1$. By induction, this prove that
$u\circ v=uv$ for an element $u$ of any degree.
By linearity, this shows that $u\circ v=uv$ 
for any $u$ and $v$ in $\ANcal$ and the property is proved
for bosons.

Adding the proper signs, the same proof shows that
$u\circ v=uv$ if  $\ANcal$ contains boson and fermion fields.

\subsubsection{Operator twisting of the algebra of fields}
The Wightman pairing $(|)_+$ on the algebra $\ANcal$ of normal
products extends to a Laplace pairing $(|)_+$ on the algebra
of fields $\AN$, that we also call the Wightman pairing. 
For scalar fields we obtain
\begin{eqnarray*}
(\phi(x)|\phi(y))_+ &=&
\int \varphi(x;\kbf)^\dagger \varphi(y;\kbf) \dd\mu(\kbf),
\end{eqnarray*}
which can again be defined as
$(\phi(x)|\phi(y))_+=\langle 0 |\phi(x)\phi(y)|0\rangle$.
The Wightman pairing for the product of Dirac fields is
\begin{eqnarray*}
(\psi(x)|\barpsi(y))_+ &=& \sum_n \varphi_>(x;n) \barphi_>(y;n)
+ \sum_{\alpha=1}^2 \int  \varphi_>(x;\kbf,\alpha) \barphi_>(y;\kbf,\alpha)
\dd\mu(\kbf),
\\
(\barpsi(x)|\psi(y))_+ &=&
\sum_{\alpha=1}^2 \int  \varphi_<(x;\kbf,\alpha) \barphi_<(y;\kbf,\alpha)
\dd\mu(\kbf),\\
(\psi(x)|\psi(y))_+ &=& 0,
\quad\quad (\barpsi(x)|\barpsi(y))_+ = 0.
\end{eqnarray*}

For scalar fields in the vacuum this gives us 
\begin{eqnarray*}
(\phi(x)|\phi(y))_+ &=& 
\int \frac{\dd\kbf}{(2\pi)^3 2\omega_k} \ee^{-ip\cdot (x-y)}
\\&=& \int \frac{\dd^4p}{(2\pi)^3} \delta(p^2-m^2) 
\theta(p^0)\ee^{-ip\cdot (x-y)}.
\end{eqnarray*}
For Dirac fields in the vacuum
\begin{eqnarray*}
(\psi(x)|\barpsi(y))_+ &=& \int \frac{\dd p}{(2\pi)^3} 
    \delta(p^2-m) \theta(p^0) (\gamma\cdot p+m) \ee^{-i p\cdot (x-y)},\\
(\barpsi(x)|\psi(y))_+ &=& - \int \frac{\dd p}{(2\pi)^3} 
   \delta(p^2-m) \theta(-p^0) (\gamma\cdot p+m) \ee^{-i p\cdot (x-y)}.
\end{eqnarray*}

The proof of Proposition~\ref{proofucircv} can be repeated to
show that
\begin{prop}
The twisted product defined by the Wightman pairing
is equal to the operator product of fields: for
any elements $u$ and $v$ of $\AN$, 
$u\circ v=uv$.
\end{prop}
Therefore, the Wightman pairing twists the algebra $\AN$ of normal
products of fields into the algebra $\AO$ of operator products
of fields.  Notice that the Dirac operator
$D=i\gamma\cdot\partial_x -m$ annihilates the Wightman pairing:
$D(\psi(x)|\barpsi(y))_+=D (\barpsi(x)|\psi(y))_+=0$.

\subsubsection{Time-ordered twisting of the algebra of fields}
We call \emph{Feynman pairing} the Laplace pairing defined by
\begin{eqnarray*}
(\phi(x)|\phi(y))_F &=& 
\theta(x^0-y^0) (\phi(x)|\phi(y))_+
+ \theta(y^0-x^0) (\phi(y)|\phi(x))_+,
\end{eqnarray*}
for scalar fields and
\begin{eqnarray*}
(\psi_\xi(x)|\barpsi_{\xi'}(y))_F &=& -(\barpsi_{\xi'}(y)|\psi_\xi(x))_F
\\
&=&
\theta(x^0-y^0) (\psi_\xi(x)|\barpsi_{\xi'}(y))_+
- \theta(y^0-x^0) (\barpsi_{\xi'}(y)|\psi_{\xi}(x))_+,\\
(\psi(x)|\psi(y))_F &=& 0, \quad\quad (\barpsi(x)|\barpsi(y))_F = 0,
\end{eqnarray*}
for Dirac fields.
The Feynman pairing is proportional to the 
Feynman propagator:
$(\psi(x)|\barpsi(y))_F = i S_F(x-y)$.
In the vacuum
\begin{eqnarray*}
(\psi(x)|\barpsi(y))_F &=& 
i\int \frac{\dd k}{(2\pi)^4} \frac{\ee^{-i k\cdot(x-y)}}{\gamma\cdot k
-m + i\epsilon}.
\end{eqnarray*}
The action of the Dirac operator on
the Feynman pairing is $D(\psi(x)|\barpsi(y))_F=i\delta(x-y)$. 

The time-ordered product satisfies the same Wick theorem
as the operator product \cite{Gross}. Thus the same proof can be used to
show
\begin{prop}
The twisted product defined by the Feynman pairing
is equal to the time-ordered product: for
any elements $u$ and $v$ of $\ANcal$, 
$u\circ v=T(uv)$.
\end{prop}

Therefore, the Feynman pairing twists the algebra $\AN$ of normal
products of fields into the algebra $\AT$ of time-ordered products
of fields.  
We saw that the twisted product is associative.
Thus, the time-ordered product of free fields
is associative. As far as we know, this property of
time-ordered products was never pointed out
explicitly.

\subsubsection{$*$-structure and real Laplace pairings}
\label{sec:lapreal}

The $*$-structure satisfies $(u\pwedge v)^*=v^*\pwedge u^*$.
A Laplace pairing is called {\em{real}}
(see \cite{Majid} p.55) when the
corresponding twisted product satisfies
$(u\circ v)^*=v^*\circ u^*$. In this section
we investigate the properties of a real Laplace pairing.
First, it can be shown that a Laplace pairing is real
if and only if $(u|v)^*=(v^*|u^*)$.

In the case of $\AOcal$, it can be checked that
the Wightman pairing $(|)_+$ is real, because the
density $\rho(\kbf)$ is real. For a neutral scalar field,
$\phi(x)^*=\phi(x)$ and 
$(\phi(x)|\phi(y))_+^*=(\phi(y)|\phi(x))_+$. Thus, the
Wightman pairing is real.
Similarly, for a Dirac field
$(\psi(x)|\barpsi(y))_+^*=(\barpsi(y)^*|\psi(x)^*)_+$.
Thus, these Wightman pairings are real and we
have $(uv)^*=v^*u^*$, which is the expected behaviour of
operator products.

The Feynman pairing $(|)_F$ corresponding to 
the time-ordered product is not real. 
However, the $*$-operation is still important because
it is related to the time-reversal symmetry
(see Section~\ref{sec:cohomtop}).

\subsubsection{Closed formulas for Wick expansion and
expectation values}
\label{sec:2.4.5}

We present here the application of the Hopf
algebra approach to the calculation of iterated
products and their vacuum expectation values.
Similar results were obtained for the bosonic
case in \cite{BrouderGroup24}.
To state these results we first need to define
the powers $\Delta^k$ of the coproduct
as $\Delta^0 a=a$,
$\Delta^1 a=\Delta a$ and
$\Delta^{k+1} a = (\id\otimes\dots\otimes \id\otimes\Delta)\Delta^k a$.
Their action is denoted by
$\Delta^k a =\sum a\i1\otimes\dots\otimes a\ii{k+1}$.

For the vacuum expectation values we have now
\begin{prop}
For $u^1,\dots,u^n$ in $\HNcal$ or $H$ we have
\begin{eqnarray}
\counit(u^1\circ\dots\circ u^n )
&=& \sum_{(u)} (-1)^{F_n} \prod_{j=2}^{n}\prod_{l=1}^{j-1}
  (u\i{j-1}^l|u\i{l}^{j}),
\label{epsilonE}
\end{eqnarray}
where the index $(u)$ means that we
sum over the required powers of the coproducts of
$u^1,\dots,u^n$ and
where
\begin{eqnarray*}
F_n &=& \sum_{i=3}^n\sum_{j=1}^{i-1}\sum_{k=2}^{i-1}\sum_{l=1}^{k-1}
 |u\i{j}^k| |u\i{l}^i|.
\end{eqnarray*}
\end{prop}
In the case of the time-ordered
product of quantum fields, the right hand side of 
equation (\ref{epsilonE}) is written as a sum of
Feynman diagrams. Our formula is also valid for the
vacuum expectation value of the operator product of
fields. An example of the application of this formula
to scalar fields was given in \cite{BrouderGroup24}.

For the Wick expansion of operator products or 
time-ordered products of fields, we
have the
\begin{prop}
For $u^1,\dots,u^n$ in $\HNcal$ or $H$ we have
\begin{eqnarray}
u^1\circ\dots\circ u^n &=&
\sum_{(u)} (-1)^{F(u)}\,
\counit(u\i1^1\circ\dots\circ u\i1^n )
u\i2^1\pwedge\dots\pwedge u\i2^n,
\label{u1circ}
\end{eqnarray}
where 
\begin{eqnarray}
F(u) &=& \sum_{k=2}^n \sum_{l=1}^{k-1} |u\i1^k||u\i2^l|.
\label{F(u)}
\end{eqnarray}
\end{prop}
In perturbative quantum field theory, this equation
is used for the calculation of the S-matrix: the
product $\circ$ is then the time-ordered product and
$u^1=\dots=u^n=\Lcal$, where $\Lcal$ is
the interaction Lagrangian of the theory.
As an example, we consider the Lagrangian 
for the $\phi^n$ theory:
$\Lcal(x)=\phi^n(x)$, where $\phi^n(x)$ denotes
the normal product of $n$ fields $\phi(x)$.
The binomial formula gives us the coproduct of $\Lcal(x)$:
\begin{eqnarray*}
\Delta \phi^n(x) &=& \sum_{k=0}^n \binom{n}{k} \phi^k(x)\otimes \phi^{n-k}(x),
\end{eqnarray*}
and equation (\ref{u1circ}) becomes, in the usual notation
\begin{eqnarray*}
T\big(\phi^{n_1}(x_1)\dots\phi^{n_m}(x_m)\big)
&=& \sum_{i_1=0}^{n_1}\dots\sum_{i_m=0}^{n_m}
\binom{n_1}{i_1}\dots\binom{n_m}{i_m}
\\&&\hspace*{-30mm}
\langle 0|T\big(\phi^{i_1}(x_1)\dots\phi^{i_m}(x_m)\big)
|0\rangle
{:}\phi^{n_1-i_1}(x_1)\dots\phi^{n_m-i_m}(x_m){:},
\end{eqnarray*}
where $T$ is the time-ordering operator and
${:}u{:}$ stands for the normal product.
This equation can be found, for example in
refs.~\cite{Epstein,Brunetti}.
Our equation (\ref{u1circ}) is more compact and
much more general: it is valid for bosons and fermions, for 
products of any elements $u$ of $H$ (and not
only of $\phi^{n}(x)$), for operator
products as well as time-ordered products.

The proof of these formulas was given in 
\cite{BrouderGroup24} for bosonic fields, so
we leave to the reader the determination of
the additional signs. However, we give the main lemmas
that lead to them.
\begin{lem}
For $u^1,\dots,u^n$ and $v^1,\dots,v^m$ in $\HNcal$ or $H$
we have
\begin{eqnarray}
(u^1\pwedge\dots\pwedge u^n|v^1\pwedge\dots\pwedge v^m)
&=&
\sum_{(u)(v)} (-1)^{F_{nm}}\prod_{i=1}^n\prod_{j=1}^m (u\i{j}^i|v\i{i}^j),
\label{u1uk}
\end{eqnarray}
where the sign $(-1)^{F_{nm}}$ is given by
\begin{eqnarray*}
F_{nm} &=& \sum_{i=1}^n \sum_{j=1}^m \sum_{(u)}
\Big(|u\i{j}^i|^2+\sum_{k=1}^i\sum_{l=1}^j
|u\i{j}^i||u\i{l}^k|\Big).
\end{eqnarray*}
\end{lem}
To obtain this equation, we used the fact that the Laplace pairing
is even, so that $|u\ii{j}^i|=|v\ii{i}^j|$. 
Two special cases are important \cite{Grosshans}: 
(i) when all $u^i$ and $v^j$ are in $\VN$ or $V$ and
are fermionic
\begin{eqnarray*}
(u^1\pwedge\dots\pwedge u^n|v^1\pwedge\dots\pwedge v^m) &=&
\delta_{m,n} (-1)^{n(n-1)/2} \det(u^i|v^j),
\end{eqnarray*}
and (ii) when all $u^i$ and $v^j$ are in $\VN$ or $V$ and
bosonic
\begin{eqnarray*}
(u^1\pwedge\dots\pwedge u^n|v^1\pwedge\dots\pwedge v^m) &=&
\delta_{m,n} \perm(u^i|v^j),
\end{eqnarray*}
where $\perm(u^i|v^j)$ is the permanent
of the matrix $(u^i|v^j)$.

To calculate iterated products recursively
we need the following identity:
\begin{lem}
For $u_1,\dots,u_n$ in $\HNcal$ or $H$,
\begin{eqnarray}
\Delta (u^1\circ\dots \circ u^n) &=&
\sum_{(u)} (-1)^{F(u)}\,
u\i1^1\circ\dots\circ u\i1^n
\otimes
u\i2^1\pwedge\dots\pwedge u\i2^n,
\label{Deltau1circ}
\end{eqnarray}
where $F(u)$ is given by equation (\ref{F(u)}).
\end{lem}

These results illustrate the power of Hopf algebra
methods to derive explicit expressions in quantum
field theory.

\section{Cohomology}
\label{sec:3}

In this section we uncover some of the deeper
mathematical structures that lie behind the twist construction of the
time-ordered product and the operator product. Principally, these are
Sweedler's Hopf algebra cohomology and
the Drinfeld twist. These new insights in turn give us new tools for
quantum field theory that will be exploited subsequently to
describe interactions (Section~\ref{sec:inter}) and non-trivial vacua
(Section~\ref{sec:vacua}). 

\subsection{Cohomology of Hopf superalgebras}
\label{sec:cohom}
 
In this section we review the basics of Sweedler's cohomology theory
of cocommutative Hopf algebras \cite{Sweedler} generalized by Majid
\cite{Majid}.
We adapt it here to the Hopf superalgebra case.


\subsubsection{Convolution product} 

Let $H$ be a Hopf superalgebra. Consider the set $L^n(H)$ of even
linear maps $H\tens\cdots\tens H\to\mathbb{C}$
on the $n$-fold tensor product of $H$. 
A linear map $\chi$ is even if
$\chi(a_1,\dots,a_n)=0$ when $|a_1|+\cdots+|a_n|$ is odd.
Let $\phi$ and $\psi$ be two even maps. We define their \emph{convolution
product} as the element in $L^n(H)$ given by
\begin{multline}
 (\phi\star\psi)(a_1,\dots, a_n)  = \\
 \sum
 (-1)^{\sum_{k=2}^{n}\sum_{l=1}^{k-1} |a_k\i1| |a_l\i2| } 
  \phi(a_1\i1,\dots, a_n\i1)
   \psi(a_1\i2,\dots, a_n\i2) . \label{eq:conv}
\end{multline}
For example, the product of $\phi,\psi\in L^1(H)$ reads simply
\[
 (\phi\star\psi)(a)=\sum \phi(a\i1)\psi(a\i2) .
\]
For $\phi,\psi\in L^2(H)$ the product is
\[
 (\phi\star\psi)(a,b)=\sum (-1)^{|b\i1| |a\i2|}
  \phi(a\i1,b\i1)\psi(a\i2,b\i2) .
\]
The convolution product makes $L^n(H)$ into an algebra. It is unital
with the unit given by
$e(a_1,\dots,a_n)=\counit(a_1)\cdots\counit(a_n)$.
Thus, a
\emph{convolution inverse} for an element $\chi\in L^n(H)$ is an
element $\chi^{-1}\in L^n(H)$ such that
$\chi\star\chi^{-1} =\chi^{-1}\star\chi =e$.
 

\subsubsection{Cochains and coboundary}
 
An \emph{$n$-cochain} is an element $\chi\in L^n(H)$ such that $\chi$
is convolution invertible and counital.
\emph{Counitality} is the property
\begin{equation}
\chi(a_1,\dots, a_{i-1}, 1, a_{i+1},\dots, a_{n}) =
\counit(a_1)\cdots\counit(a_{i-1})\counit(a_{i+1})\cdots\counit(a_{n}),
\label{eq:counital}
\end{equation}
for all $i\in\{1,\dots,n\}$.
We denote by $C^n(H)$ the set of $n$-cochains on $H$.
The set $C^n(H)$ forms a group with the convolution product.
The unit element is the cochain 
$e(a_1,\dots,a_n)=\counit(a_1)\cdots\counit(a_n)$.
 
For $i=0,\dots n+1$ consider the maps $\partial^n_i:C^n(H)\to
C^{n+1}(H)$ defined by
\begin{gather*}
 (\partial^n_i\chi)(a_1, \dots, a_{n+1})
 =\chi(a_1,\dots,a_{i}a_{i+1},\dots,a_{n+1})\quad\forall i\in\{1,\dots,n\},\\
 (\partial^n_0\chi)(a_1,\cdots,a_{n+1})
 =\counit(a_1)\chi(a_2,\dots,a_{n+1}),\\
 (\partial^n_{n+1}\chi)(a_1,\dots,a_{n+1})
 =\chi(a_1,\dots,a_{n})\counit(a_{n+1}).
\end{gather*}
The map $\partial^n:C^n(H)\to C^{n+1}(H)$ defined by
\begin{equation}
 \partial^n\chi=(\partial^n_0\chi)\star(\partial^n_2\chi)
  \star\cdots\star(\partial^n_1\chi^{-1})\star(\partial^n_3\chi^{-1})
  \star\cdots
 \label{eq:defcob}
\end{equation}
is called the \emph{coboundary} map.\footnote{Note
that the group operation appearing in the definition of the coboundary
map is not the vector space addition as
e.g.\ in Hochschild cohomology. Nevertheless, in order to give rise to
a cohomology the group operation has to be abelian, see below.}
For example,
\begin{eqnarray}
\partial^1 \chi(a,b) &=&  \sum \chi(a\i1)\chi(b\i1)
             \chi^{-1}(a\i2 b\i2), \label{eq:1cob} \\
\partial^2 \chi(a,b,c) &=&  \sum 
  (-1)^{|a\i1||a\i2|+ |c\i2||b\i3|+ |b\i1||b\i2|}\nonumber \\ 
&& (-1)^{|b\i1||b\i3|+|b\i1||b\i4|+ |b\i2||b\i4|+ |b\i3||b\i4|}\nonumber \\
&& \chi(b\i1,c\i1)\chi(a\i1,b\i2 c\i2)\chi^{-1}(a\i2 b\i3, c\i3)
\chi^{-1}(a\i3,b\i4).
 \label{eq:2cob} 
\end{eqnarray}
We usually write just $\partial$ instead of $\partial^n$.

\subsubsection{Cohomology groups}

An $n$-cochain $\chi$ with the property $\partial\chi=e$ is called an
$n$-\emph{cocycle}. The cocycles from a subset $Z^n(H)$ of
$C^n(H)$.
Explicitly, the cocycle condition for a 1-cochain comes out as
\begin{equation}
 \chi(a)\chi(b)=\chi(a b),
\end{equation}
while the 2-cocycle condition can be written as
\begin{multline}
 \sum (-1)^{|b\i2| |c\i1|}\chi(b\i1, c\i1)\chi(a,b\i2 c\i2) = \\
 \sum (-1)^{|a\i2| |b\i1|}
\chi(a\i1,b\i1) \chi(a\i2 b\i2,c) .
 \label{eq:cocycleid}
\end{multline}
An $n$-cochain $\chi$ that arises from an $(n-1)$-cochain $\xi$
as $\chi=\partial\xi$ is called an $n$-\emph{coboundary}. The
coboundaries also form a subset $B^n(H)$ of $C^n(H)$.

Now assume that $H$ is graded cocommutative. The convolution product is then
commutative and $C^n(H)$ is an Abelian group. Furthermore $\partial$
becomes a group homomorphism and both $Z^n(H)$ and $B^n(H)$ become
groups. Moreover, we then have 
\[
 \partial\partial\xi=e,
\]
so that an $n$-coboundary is in particular an $n$-cocycle.
Thus $B^n(H)$ is a subgroup of $Z^n(H)$
and we can form the quotient group $H^n(H)=Z^n(H)/B^n(H)$.
This is called the $n$th cohomology group of $H$.\footnote{Of course
only in this case of graded cocommutativity the word
``cohomology'' is fully justified. Indeed this is the only situation
of interest in the present paper. However, important elements of the
cohomology remain applicable in the case of non
(graded) cocommutative Hopf algebras.}


\subsection{Drinfeld twist}
\label{sec:twist}

In this section we review basic properties of Drinfeld twists due to
Drinfeld \cite{Drinfeld} and Sweedler \cite{Sweedler}. We present a
version adapted to Hopf superalgebras.

We first recall the notions of comodule (representation) and comodule
superalgebra.
A (left) \emph{comodule} of a Hopf algebra $H$ is a vector space
$A$ together with a linear map $\beta:A\rightarrow H\otimes A$ such that
$(\id\tens\beta)\circ\beta = (\cop\tens\id)\circ\beta$ and
$(\cou\tens\id)\circ\beta = \id$. $\beta$ is called a
\emph{coaction}. For coactions we also use a modified version of
Sweedler's notation with the component in the comodule underlined,
$\beta (a)=\sum a\i1\otimes a\iu2$.

Consider a comodule $A$ of $H$ that is at the same time a
superalgebra. It is called a \emph{comodule superalgebra} of $H$ if
product (denoted by $\cdot$) and comodule structure satisfy the
following compatibility condition:
\begin{eqnarray*}
 \sum (a \cdot b)\i1\otimes (a \cdot b)\iu2 = \sum (-1)^{|b\i1| |a\iu2|}
  a\i1 b\i1\otimes a\iu2 \cdot b\iu2 .
\end{eqnarray*}

\subsubsection{Twisting Hopf superalgebras and comodules}

Let $H$ be a Hopf superalgebra and $\chi\in Z^2(H)$ a 2-cocycle on
$H$. There is a new Hopf superalgebra $H_\chi$, the \emph{twist} of
$H$ by $\chi$. $H_\chi$ has the same unit, counit and coproduct as $H$ but
a different product and antipode. The new product is given by
\begin{eqnarray}
 a\bullet b &=&\sum (-1)^{|b\i1| (|a\i2|+|a\i3|) + |b\i2| |a\i3|}
\nonumber
\\&&\hspace*{5mm}
 \chi(a\i1,b\i1) a\i2 b\i2 \chi^{-1}(a\i3,b\i3) .
\label{eq:dtwist}
\end{eqnarray}

It turns out that a twist can be applied not only to the Hopf
superalgebra itself but also to its comodules.\footnote{The twist
  gives rise to an equivalence of 
the monoidal categories of comodules of $H$ and $H_\chi$ \cite{Drinfeld}. 
This is explained in detail in \cite{OecklPhD}.}
If a comodule $A$ is
a comodule superalgebra the twist affects its superalgebra
structure. $A$ is twisted into a comodule superalgebra $A_\chi$ of
$H_\chi$ with the new associative product $\circ$ defined by
\begin{equation}
 a\circ b= \sum (-1)^{|b\i1| |a\iu2|}
 \chi(a\i1,b\i1) a\iu2 \cdot b\iu2 .
 \label{eq:twistcomalg}
\end{equation}
 
If $H$ is graded cocommutative, its product remains unmodified under a
twist and $H_\chi$ is the same as $H$. However, this is not true for a
comodule superalgebra $A$ of $H$. In general $A_\chi$ is different from
$A$ even in the graded cocommutative case. Indeed, the difference
between twisted comodule superalgebras is related to the cohomology of
$H$ as follows:

\begin{prop}
\label{prop:twistcohom}
Let $H$ be a graded cocommutative Hopf superalgebra, $A$ a left
$H$-comodule superalgebra and $\eta,\chi\in Z^2(H)$.
If $\eta$ and $\chi$ are cohomologous in the sense of
$\eta=\partial\rho\star \chi$ for $\rho\in C^1(H)$, then $A_\eta$ and
$A_\chi$ are isomorphic as comodule superalgebras. An isomorphism
$T:A_\eta\to A_\chi$ is explicitly given by
$T(a)=\sum \rho(a\i1) a\iu2$.

If $A=H$ with the coaction given by the coproduct, the converse is
also true. That is, if $A_\eta$ and $A_\chi$ are isomorphic as
comodule superalgebras then
$\eta$ and $\chi$ are cohomologous.
\end{prop}
A proof based on \cite{Majid} can be found in Appendix~\ref{sec:proofs}.

\subsubsection{Twisting and $*$-structure}

Suppose that $H$ is a Hopf $*$-superalgebra in the sense of
Appendix~\ref{sec:starsuper} and $A$ a graded left comodule of
$H$ equipped with an involution $*:A\to A$. Then we call $A$ a
$*$-comodule if the coaction satisfies
\begin{equation}
 \sum (a^*)\i1\tens (a^*)\iu2
 = \sum (-1)^{|a\i1| |a\iu2|} (a\i1)^* \tens (a\iu2)^* .
\label{eq:starcom}
\end{equation}
In the same way we can define a $*$-comodule
superalgebra $A$. In this case we can furthermore investigate under
which circumstances a 2-cocycle $\chi$ gives rise to a twisted
comodule superalgebra $A_\chi$ which is again a
$*$-superalgebra. Indeed it is straightforward to verify that a
sufficient condition on $\chi$ for this to happen is
\begin{equation}
 \chi(a^*, b^*)=\overline{\chi(b, a)} .
\label{eq:real}
\end{equation}
A 2-cocycle satisfying this property we call \emph{real}. Our
definition is inspired by the analogous definition for coquasitriangular
structures in the literature \cite[Definition~2.2.8]{Majid}.
It extends the definition for real Laplace pairing given in
Section~\ref{sec:lapreal}.


\subsection{Cohomology in quantum field theory}

We are now ready to interpret and extend the results of
Section~\ref{sec:2} from a cohomological point of view.

\subsubsection{The twisted field algebras as Drinfeld twists}
\label{sec:twistfield}

Recall from Section~\ref{sec:salgfields} that the field operators
with the normal product form a Hopf superalgebra $H$. Forgetting the
coproduct and counit we also denoted this superalgebra by $A_N$.
Then we discovered in Section~\ref{sec:opto} that for certain Laplace
pairings on this Hopf
algebra we obtain new twisted algebras $A_O$, $A_T$ using
(\ref{defcircle}) which
reproduce either the operator or the time-ordered product.

Armed with the tools of Hopf algebra cohomology
(Section~\ref{sec:cohom}) and the Drinfeld twist
(Section~\ref{sec:twist}) we see more clearly what is going
on. Namely, the Laplace pairings give rise to Drinfeld twists of
$A_N$ as a comodule superalgebra of $H$.
Let us clearly explain this step by step.

Firstly, a Laplace pairing (in the graded cocommutative case)
is in particular a 2-cocycle.
\begin{lem}
\label{lem:lapcoc}
Let $H$ be a graded cocommutative Hopf superalgebra. Then a Laplace
pairing, i.e.\ a map $\chi:H\tens H\to \C$ with the properties
(\ref{eq:pair1}), (\ref{eq:pair2}) and
(\ref{eq:pair3}) is a 2-cocycle.
\end{lem}
\begin{proof}
Equation (\ref{eq:pair3}) is the counitality property
(\ref{eq:counital}). Let $\eta$ be the linear map $H\tens
H\to \C$ defined by $\eta(a, b)\defeq \chi(\antip(a), b)$. It is
elementary to check that $\eta$ is the convolution inverse of
$\chi$. Thus, $\chi$ is a 2-cochain. Finally, using graded
cocommutativity the cocycle condition (\ref{eq:cocycleid}) readily
follows from equations (\ref{eq:pair1}) and (\ref{eq:pair2}).
\end{proof}

Secondly, $A_N$ is a comodule superalgebra of $H$. Indeed, any
Hopf superalgebra has a comodule superalgebra which
is just a copy of itself. The coaction to be taken is the coproduct,
i.e.\ $\sum a\i1\tens a\iu2\defeq \sum a\i1\tens a\i2$ in the notation
introduced above. A Laplace pairing $\chi$ on $H$, being a 2-cocycle,
gives rise to a Drinfeld twist of $H$ according to
(\ref{eq:dtwist}). Since $H$ is graded
cocommutative the twisted Hopf superalgebra $H_\chi$ is identical to
the untwisted one. $\chi$ also gives rise to an induced twist of the
comodule superalgebra $A_N$ according to (\ref{eq:twistcomalg}). Since
the coaction is given by the coproduct we recover the initial twist
formula (\ref{defcircle}) with a new interpretation. This also
explains why the twisted superalgebras $A_O$ and $A_T$ are no longer
Hopf algebras: $A_N$ was not considered a Hopf algebra from the
beginning, despite the ``accident'' $A_N=H$.\footnote{One might
envision applications to quantum field theory where $H$ is different
from $A_N$, but this is beyond the scope of the present paper.}

Recall that apart from the superalgebra structure we use one more piece
of the Hopf algebra structure of $H$ on $A_N$. This is the map
$\cou:A_N\to\C$ which is the counit on $H$. As was shown in
Proposition~\ref{prop:vev} it plays the role of the vacuum expectation
value. The twisted superalgebras $A_O$ and $A_T$ inherit the map
$\cou$ without change.\footnote{This is perfectly justified
from the Drinfeld twist
point of view. Morally speaking, the twist does not affect comodule
structures and comodule maps as such but the tensor product of
comodules (and hence a comodule algebra structure as its definition
involves a tensor product of comodules). The deeper meaning of this lies
in the fact that the twist gives rise to a monoidal equivalence of
comodule categories. This equivalence is mediated by a functor that
transforms objects and morphisms trivially and only tensor products
non-trivially. See Section~2 of \cite{OecklPhD} for a more explicit
exposition of these facts.}
$\cou$ continues to play the role of the vacuum
expectation value. Only equipped with $\cou$
carry the superalgebras the full information of quantum field
theory. Superalgebras such as $A_N, A_O, A_T$, which carry
the additional structure of a linear function $A\to\C$ are called
\emph{augmented} superalgebras.

For the $*$-structure we remark that $A_N$ and $H$ could in principle
have turned out to have different $*$-structures. What is important is
only that $A_N$ is a $*$-comodule superalgebra of $H$ in the sense of
(\ref{eq:starcom}). The results of Section~\ref{sec:2}, however, imply that
putting the same $*$-structure on $H$ and $A_N$ (namely that of the
$\dagger$ operation in $A_N$) leads to consistent results.

\subsubsection{Cohomology of $\Sym(V)$}

We turn to cohomological aspects of the relevant Hopf superalgebra
$H$ of field operators. Recall that $H$ has the structure of the
graded symmetric Hopf superalgebra $\Sym(V)$, where $V$ is the space
of field operators.
The cohomology groups of $\Sym(V)$ (in the bosonic case) are discussed
in Appendix~\ref{sec:cohomgrp}. More relevant for the application to
quantum field
theory, are the following results about the structure of cochains. The
proofs of the Lemmas are elaborated in Appendix~\ref{sec:proofs}.

Let us write
$V= V_0 \oplus V_1$ for $V_0$ the even (bosonic)
and $V_1$ the odd (fermionic) part of $V$. To ease notation we simply
write $C^n$,
$Z^n$ and $B^n$ for cochains, cocycles and coboundaries of $\Sym(V)$.
Denote by $N^1$ the set of 1-cochains of $\Sym(V)$ that
vanish on the subspace $V\subset \Sym(V)$. As is seen immediately they
form a subgroup of $C^1$.

\begin{lem}
\label{lem:1coch}
$C^1$ is equal to the direct product $Z^1\times N^1$ of its subgroups.
This implies that $\partial^1: N^1\to B^2$ is
invertible, i.e.\ is an isomorphism of groups.
\end{lem}

We call a 2-cochain $\chi$ \emph{symmetric} if it satisfies the
property
\begin{equation}
\chi(b,a)=(-1)^{|a| |b|}\chi(a,b)
\label{eq:cocsym}
\end{equation}
for all $a,b\in
\Sym(V)$. One easily checks that the symmetric 2-cochains form a subgroup
$C^2_\mathrm{sym}$ of $C^2$.

\begin{lem}
\label{lem:cobsym}
 $B^2=Z^2_\mathrm{sym}$. That is, the 2-coboundaries are precisely the
 symmetric 2-cocycles.
\end{lem}

By Lemma~\ref{lem:lapcoc} a Laplace pairing as defined in
Section~\ref{sec:deflap} is in particular a 2-cocycle. Furthermore, the
convolution product of Laplace pairings is again a Laplace
pairing. Thus, they form a subgroup of $Z^2$ which we will call
$R^2$.

The introduction of something like
``antisymmetric'' 2-cochains is less straightforward. We limit
ourselves here to Laplace pairings. We call a Laplace pairing
\emph{antisymmetric} if it satisfies the property
\begin{equation}
\chi(w,v)=-(-1)^{|v| |w|}\chi(v,w)
\label{eq:lapasym}
\end{equation}
for all $v,w\in V$. The antisymmetric Laplace pairings
form a subgroup $R^2_\mathrm{asym}$ of $R^2$.

\begin{lem}
\label{lem:2cocycle}
 $Z^2$ is equal to the direct product $B^2\times R^2_\mathrm{asym}$ of
 its subgroups.
\end{lem}

From Lemma~\ref{lem:1coch} and Lemma~\ref{lem:2cocycle} it follows
that the first two cohomology groups are given by $H^1=N^1$ and
$H^2=R^2_\mathrm{asym}$, but we shall not need these results here.


\subsubsection{The operator product}

We saw in Section~\ref{sec:opto} that the operator product of quantum
field theory, i.e.\ the product on $A_O$, is induced by a twist of
$A_N$ with the Wightman pairing $(|)_+$ on $H$. The fact that $A_O$
is not commutative and thus not isomorphic to $A_N$ is nicely
reflected in the cohomology. Namely, the 2-cocycle $(|)_+$ is
not symmetric in
the sense of (\ref{eq:cocsym}) and thus by Lemma~\ref{lem:cobsym} it
is not a 2-coboundary. Hence 
by Proposition~\ref{prop:twistcohom} (take $\eta=e$, the unit cochain,
and $\chi=(|)_+$)
the algebras $A_N$ and $A_O$
cannot be isomorphic.

Since the Wightman pairing $(|)_+$ is real in the sense of (\ref{eq:real})
the superalgebra $A_O$ is a $*$-superalgebra as is $A_N$. This was
already remarked in Section~\ref{sec:lapreal}.


\subsubsection{The time-ordered product}
\label{sec:cohomtop}

According to Section~\ref{sec:opto} the time-ordered product is obtained
from the Feynman pairing $(|)_F$, twisting the superalgebra $A_N$ into
the superalgebra $A_T$. As is well known these superalgebras $A_N$ and
$A_T$ are isomorphic as superalgebras.
From the cohomological point of
view this emerges as follows. The Feynman pairing $(|)_F$ is symmetric in
the sense of (\ref{eq:cocsym}) and
thus by Lemma~\ref{lem:cobsym} it is a 2-coboundary. Hence by
Proposition~\ref{prop:twistcohom} the algebras $A_N$ and $A_T$ are
isomorphic.

However, this does \emph{not} mean that the
``deformation'' of $A_N$ into $A_T$ is trivial from the point of view
of quantum field theory. Recall that the information of quantum
field theory is not contained in $A_T$ alone, but crucially requires a
map $\cou:A_T\to\C$ describing the vacuum expectation
value. That is, we are dealing with augmented superalgebras and as
such $(A_N,\cou)$ and $(A_T,\cou)$ are not isomorphic. In particular,
for an isomorphism $T:A_N\to A_T$ the composition $\cou\circ T$ is
different from $\cou$.

Proposition~\ref{prop:twistcohom}  (take $\eta=e$
and $\chi=(|)_F$) not only tells us that $A_N$ and
$A_T$ are isomorphic but even provides us with one explicit isomorphism
$T:A_N\to A_T$ for each 1-cochain $\rho$ that satisfies
$\chi=(\partial\rho)^{-1}$. The choice of such 1-cochains $\rho$ is
parametrized by a 1-cocycle.
We can select a unique 1-cochain $\rho$ by imposing suitable
conditions on the
associated isomorphism $T$. A natural choice is to demand $T$ to act
identically on $V$, i.e.\ $T(v)=v$ for all $v\in V$. This is motivated
by giving $T$ the role of a time-ordering operation. Since $T$
is given according to Proposition~\ref{prop:twistcohom} by
$T(a)=\sum\rho(a\i1) a\iu2$ this would imply $\rho(v)=0$ for all $v\in
V$, i.e.\ that $\rho\in N^1$. Lemma~\ref{lem:1coch} implies
that we can indeed choose $\rho$ to lie in $N^1$ and
furthermore, that this determines $\rho$ uniquely. This gives rise to
the following result.

\begin{prop}
\label{prop:tord}
Let $\chi$ be a 2-coboundary on $H$. Denote by $\circ$ the induced twisted
product on the twisted comodule superalgebra $A_T$. There exists a
unique 1-cochain $\rho$ such that $(\partial\rho)^{-1}=\chi$ and
$\rho(v)=0$ for all
$v\in V$. The superalgebra isomorphism $T:A_N\to A_T$ given by
$T(a)=\sum\rho(a\i1) a\iu2$ satisfies
\begin{equation}
T(v_1\pwedge\cdots\pwedge v_{n}) = v_1\circ\cdots\circ v_{n},
\label{eq:Tu}
\end{equation}
for $v_1,\dots,v_n\in V$.
\end{prop}
\begin{proof}
As already mentioned $\rho(v)=0$ for $v\in V$ implies $T(v)=v$. Thus
$T$ being an isomorphism implies $T(v_1\pwedge\cdots\pwedge
v_{n})=T(v_1)\circ\cdots\circ T(v_n)= v_1\circ\cdots\circ v_n$. This
is all that remained to be shown.
\end{proof}
Note that we have formulated the Proposition in a slightly more
general way than required, by replacing the Feynman pairing with a general
2-coboundary.
As desired, equation (\ref{eq:Tu}) can be interpreted as a realization of
the time-ordering operation of quantum field theory as a superalgebra
isomorphism between $A_N$ and $A_T$.

By definition of $T$ the 1-cochain $\rho$ has the property
$\rho=\cou\circ T$. This implies that a vacuum expectation value can
be expressed directly in terms of $\rho$.
\begin{cor}
\begin{eqnarray}
\langle 0|\phi(x_1)\circ\cdots\circ\phi(x_n) |0\rangle
& = & \cou(\phi(x_1)\circ\cdots\circ\phi(x_n)) \nonumber\\
& = & \cou(T(\phi(x_1)\pwedge\cdots\pwedge\phi(x_n))) \nonumber\\
& = & \rho(\phi(x_1)\pwedge\cdots\pwedge\phi(x_n)) .\label{eq:freen}
\end{eqnarray}
\end{cor}
In this sense, $\rho$ encodes directly the free $n$-point functions.

While we were so far only concerned with the definition of $\rho$ we
turn now to its computation. Due to Lemma~\ref{lem:1coch}, $\rho\in
N^1(H)$ is determined completely by $(|)_F=(\partial\rho)^{-1}$ as
$\partial^1$ is invertible on $N^1(H)$. Indeed we can use (\ref{eq:1cob}) 
for a recursive definition of $\rho$. Namely, set $\rho(1)=1$,
$\rho(v)=0$ and $\rho(v\pwedge w)=(v|w)_F$. Then define $\rho$
recursively on subspaces of $A_N$ of increasing degree by
\[
 \rho(a\pwedge b)=\sum (a\i1|b\i1)_F \rho(a\iu2) \rho(b\iu2) .
\]

As already mentioned in Section~\ref{sec:lapreal} the Feynman pairing
$(|)_F$ is
not real and $A_T$ is indeed not a $*$-superalgebra. Nevertheless, the
involution $*$ can be combined with the time-ordering map $T$ in an
interesting way. Namely, consider the map $T^*(a)\defeq(T(a^*))^*$ for
$a\in A_N$.
This anti-time-ordering operator
was first considered by Dyson \cite{Dyson51I}
and plays an important role in 
non-equilibrium quantum field theory
\cite{SchwingerJMP,Keldysh,Chou}.
Our definition of the anti-time-ordering operator 
follows ref. \cite{Dutsch}.
It yields (here for fermionic fields $\psi$)
\begin{eqnarray*}
T^*\big(\psi(x)\pwedge\psi^\dagger(y)\big)
&=&
\theta(y^0-x^0)\psi(x)\psi^\dagger(y)
-\theta(x^0-y^0) \psi^\dagger(y)\psi(x).
\end{eqnarray*}
In other words, $T^*$ orders the fields 
by decreasing times from right to left.
Defining the 1-cochain $\rho^*=\cou\circ T^*$ we see that the
anti-time-ordered product is a twisted product via the Laplace pairing
$(|)_{T^*}=(\partial \rho^*)^{-1}$. More explicitly (again for the
example of fermionic fields) this Laplace pairing is determined by
$(\psi(x)|\psi^\dagger(y))_{T^*}=
\overline{(\psi(y)|\psi^\dagger(x))_F}$. The map $T^*$ thus becomes an
algebra isomorphism $T^*:A_N\to A_{T^*}$, with $A_{T^*}$ the algebra
of field operators with the anti-time-ordered product.

In non-relativistic quantum theory, time-reversal
symmetry is implemented by complex conjugation
\cite{Wigner,Sachs}.
In relativistic quantum field theory, the time-reversal operator
$\Theta$ acts on fermion fields by 
$\Theta \big(\psi(x^0,\xbf)\big)=i\gamma^1\gamma^3\psi(-x^0,\xbf)$,
which does not involve the $*$-structure.
The time-reversal operator relates the
time-orderings by $T\big(\Theta(a)\big)=\Theta\big(T^*(a)\big)$
for $a\in A_N$.

\section{Interactions}

\label{sec:inter}

Up to now we have exclusively dealt with free quantum field theory. In
this section we extend our treatment to interacting fields. On the
one hand we will show how our approach to quantum field theory links up
with standard perturbation theory. On the other hand we will discuss
implications for interacting quantum field theory in general, and
possible connections to non-perturbative approaches.

\subsection{Standard perturbation theory} 

Introducing interactions in the standard perturbative way is
straightforward, given the free $n$-point functions.
Let us generically denote field operators by $\phi(x)$,
leaving out internal indices.

Following the usual perturbation theory we write the action as $S=S_0+
\lambda S_\mathrm{int}$, where $S_0$ is the kinetic term and $\lambda$ the
coupling constant. Following a path integral notation
the interacting $n$-point functions are given by
\begin{eqnarray}
 \langle 0 |T(\phi_{\mathrm{int}}(x_1), \dots, \phi_{\mathrm{int}}(x_n))
 |0\rangle
 &=&  \frac{\int \mathcal{D} \phi\, \phi(x_1)\cdots\phi(x_n)
  e^{i S_0+ i \lambda S_\mathrm{int}}}
 {\int \mathcal{D} \phi\, e^{i S_0+ i \lambda S_\mathrm{int}}}
 \nonumber\\
  &=&  \frac{\sum_k \frac{1}{k!} (i \lambda)^k
 \langle 0 | T(\phi(x_1)\pwedge\cdots\pwedge\phi(x_n)\pwedge
 S_\mathrm{int}^{\pwedge k}) | 0\rangle}
 {\sum_k \frac{1}{k!} (i \lambda)^k
 \langle 0 | T(S_\mathrm{int}^{\pwedge k}) | 0\rangle}
 \nonumber\\
  &=&  \frac{\sum_k \frac{1}{k!} (i \lambda)^k
 \rho(\phi(x_1)\pwedge\cdots\pwedge\phi(x_n)\pwedge
  S_\mathrm{int}^{\pwedge k})}
 {\sum_k \frac{1}{k!} (i \lambda)^k
 \rho(S_\mathrm{int}^{\pwedge k})} .
\label{Green}
\end{eqnarray}
Here, $S_\mathrm{int}^{\pwedge k}$ denotes the $k$-fold normal product
of $S_\mathrm{int}$ with itself and $S_\mathrm{int}^{\pwedge
0}=1$. Alternatively, in terms of an S-matrix
\[
 \mathcal{S}=\sum_{k=0}^\infty  \frac{1}{k!} (i \lambda)^k
 (T(S_\mathrm{int}))^{\circ k}
\]
we obtain
\begin{equation}
\langle 0 |T(\phi_{\mathrm{int}}(x_1), \dots, \phi_{\mathrm{int}}(x_n))
 |0\rangle
=\frac{\cou(\phi(x_1)\circ\cdots\circ\phi(x_n)
 \circ\mathcal{S})}{\cou(\mathcal{S})} ,
\label{Green:Hopf}
\end{equation}
where $\circ$ denotes the time-ordered product.

We emphasize, that the difference to conventional approaches is not
only notational.
While the path integral
method is capable of the evaluation of eqn. (\ref{Green}), it is usually
evaluated by recursive means using functional derivatives. Such algorithms
were demonstrated to be computationally ineffective \cite{AF1,AF2}. In such a
recursion, terms occur which cancel out during further steps of evaluation
in the recursion. Since these terms may even contain divergent integrals
which need to be renormalized, this can cause confusion and wastes labor.
In contrast to this finding, the Hopf algebraic version
eqn. (\ref{Green:Hopf})
can be evaluated directly using the formulas provided in
Section~\ref{sec:2.4.5}. The remarkable fact behind these formulas is
that they are
explicit and efficient for any order. It can be shown that they yield
exactly as many terms as are potentially non zero. Especially one may
note that
no cancellations take place, as long as no further symmetries are
encountered in
the pairings involved in the employed twists. The formulas
(\ref{epsilonE}) and (\ref{u1circ}) prove to be algorithmically
optimal in this sense.

\subsection{Beyond perturbation theory} 

We turn now to general considerations of quantum field theory beyond
any perturbation theory. We consider the same field operators and
thus the same Hopf superalgebra $H$ and superalgebra $A_N$ as before.
Recall that all the information of a quantum field theory (interacting
or not) is encoded in the $n$-point functions. We write generically
\[
 \rho_\mathrm{int}(\phi(x_1)\pwedge\cdots\pwedge\phi(x_n)) \defeq
 \langle 0 |T(\phi_{\mathrm{int}}(x_1), \dots, \phi_{\mathrm{int}}(x_n))
 |0\rangle .
\]
From this point of view, the set of $n$-point functions is nothing but
a 1-cochain $\rho_\mathrm{int}$ on $H$ (since $\langle 0 |1 |0\rangle=1$).

For the free theory we saw in
Section~\ref{sec:cohomtop} that the Feynman pairing $(|)_F$ leads to a
1-cochain $\rho$ that encodes directly the $n$-point functions
(\ref{eq:freen}). $\rho$ was the 1-cochain in $N^1(H)$ determined
by the property $(|)_F=(\partial\rho)^{-1}$. Indeed, we can turn this
argument round:

\begin{prop}
Let $\rho\in N^1(\Sym(V))$, then $\chi=(\partial\rho)^{-1}$ is a
2-cocycle which induces a twisted product $\circ$
in $\Sym(V)$ as a comodule superalgebra over
itself with the property
\[
 \rho(v_1\pwedge\cdots\pwedge v_n)=\cou(v_1\circ\cdots\circ v_n) ,
\]
for $v_1,\dots,v_n\in V$.
\end{prop}
\begin{proof}
By Proposition~\ref{prop:twistcohom} $T(a)=\sum \rho(a\i1) a\iu2$ is a
superalgebra isomorphism between the original and the twisted
superalgebra. The statement is then obtained by applying the counit to
(\ref{eq:Tu}) with the proof as in Proposition~\ref{prop:tord}.
\end{proof}

This means, given an arbitrary set of $n$-point functions
$\rho_\mathrm{int}$
satisfying $\rho_\mathrm{int}(1)=\langle 0| 1 |0\rangle=1$ and
$\rho_\mathrm{int}(\phi(x))=\langle 0| \phi_\mathrm{int}(x)
|0\rangle=0$, we can construct a twisted product $\circ$ which
recovers these $n$-point functions.
This product is thus the
time-ordered product of interacting fields. The 2-cocycle inducing
this product is simply
$\chi_\mathrm{int}\defeq(\partial\rho_\mathrm{int})^{-1}$.
We can view this as a kind of (algebraic) reconstruction result in the
spirit of Streater and Wightman \cite{StWi:pct}, although for the
time-ordered and not the operator product.

Furthermore, according to Proposition~\ref{prop:twistcohom} and
analogous to Section~\ref{sec:cohomtop} we obtain
an isomorphism of superalgebras $T_\mathrm{int}:A_N\to
A_{T,\mathrm{int}}$ via $T_\mathrm{int}(a)\defeq\sum\rho_\mathrm{int}(a\i1)
a\iu2$. This isomorphism might be viewed as an interacting
time-ordering, i.e.\ it takes care at the same time of the interaction
and the time-ordering. Thus, we may write
\[
  \langle 0 |T(\phi_{\mathrm{int}}(x_1), \dots, \phi_{\mathrm{int}}(x_n))
 |0\rangle =
 \cou(T_\mathrm{int}(\phi(x_1)\pwedge\cdots\pwedge\phi(x_n))) .
\]

What we have thus shown is that not only a free quantum field theory
can be completely encoded in a 2-cocycle on $H$, but \emph{any}
quantum field theory defined through polynomial $n$-point functions
can be thus encoded (provided its 1-point
functions vanish). What is more, whether the theory is free or not
corresponds to a simple property of the 2-cocycle.
We define \emph{free} here to mean that $n$-point functions factorize
into 2-point functions according to (\ref{epsilonE}).

\begin{prop}
Let a quantum field theory be given in terms of $H=\Sym(V)$ and a 2-cocycle
$\chi$ on $H$ inducing the interacting time-ordered product. Then the
theory is free if and only if $\chi$ is a Laplace pairing.
\end{prop}

The proof is straightforward now. If $\chi$ is a Laplace pairing the
$n$-point functions are determined by the 2-point functions according
to formula (\ref{epsilonE}) and thus the theory is
free. Conversely, if the theory is free formula (\ref{epsilonE})
holds and we can thus construct a Laplace pairing that reproduces the
$n$-point functions. Since these determine $\chi$ uniquely it must be
identical to the constructed Laplace pairing.
 
\section{Non-trivial vacua}
\label{sec:vacua}

In this section, we illustrate again the computational power of Hopf
algebras by solving a problem of quantum chemistry.

A {\emph{state}} is a linear map $\rho$ from $\ANcal$ to
$\mathbb{C}$ such that $\rho(1)=1$
and $\rho(u^* u)\ge 0$ for any element
$u$ of $\ANcal$ \cite{Bratteli1} (recall
that $u^* u$ is the operator product of 
$u^*$ and $u$).
The {\emph{pure states}} are states of the form
$\rho(u)=\langle\psi|u|\psi\rangle$,
where $|\psi\rangle$ is a vector of the
Fock space $\Fcal$, and
the states that are not pure are called
{\emph{mixed states}}. They can be
written as a weighted sum of pure states.
Physically relevant pure states are such
that $\langle\psi|u|\psi\rangle=0$ if
$u$ contains an odd number of Dirac fields.
Thus we consider states $\rho(u)$ 
which are zero if $u$ contains an odd number
of Dirac fields. Since $\rho$ is linear, even and
$\rho(1)=1$, a state is 1-cochain.

In reference \cite{Kutzelnigg},
Kutzelnigg and Mukherjee study the following problem
of quantum chemistry.
Assume that a quantum system is described by a state $\rho$,
is it possible to define normal products adapted to
$\rho$? The usual normal products are adapted to
the vacuum because $\epsilon(u)=\langle 0|u|0\rangle$
is zero if $u$ has no scalar part (i.e. no part
proportional to 1).
To adapt a normal product to a state $\rho$, we
start from an element $u\in\ANcal$ and we want 
to find an element $\tildu$ such that  
$\rho(\tildu)=\epsilon(u)$, so that the
state $\rho$ can be considered as the new vacuum of the system.
Kutzelnigg and Mukherjee investigated this problem
for pure states and they solved it for 
elements $u$ which are the normal product
of a small number of creation and annihilation 
operators. The Hopf algebra
methods will enable us to solve it for general
states and to give formulas that are valid for
any $u\in\ANcal$.

The solution of this problem is quite simple.
We just have to define
\begin{eqnarray}
\tildu &=& \sum \rho^{-1}(u\i1) u\i2,
\label{tildeu}
\end{eqnarray}
because
$\rho(\tildu) = \sum \rho^{-1}(u\i1) \rho(u\i2) =
(\rho^{-1}\star \rho)(u) = \counit(u)$.
Notice that formula (\ref{tildeu}) is another instance
of a T-operator, where $\rho$ is replaced by $\rho^{-1}$.

Let us give a few examples. In this section we 
consider Dirac fields, which are relevant for this
type of applications.
The convolution inverse $\rho^{-1}$ is even because $\rho$ is
even. It can be computed recursively
by (see reference \cite{MilnorMoore}, p. 259)
\begin{eqnarray*}
\rho^{-1}(1) &=& 1,\\
\rho^{-1}(u) &=& -\rho(u) - {\sum}' \rho^{-1}(u\i1) \rho(u\i2),
\end{eqnarray*}
where $\sum' u\i1\otimes u\i2 \defeq \Delta u - 1\otimes u - u\otimes 1$
for $u\in \ANcal$ and $u$ contains the product of
two or more creation or annihilation operators.
The first examples for Dirac fields are
\begin{eqnarray*}
\rho^{-1}(a\pwedge b) &=& -\rho(a\pwedge b), \\
\rho^{-1}(a\pwedge b\pwedge c\pwedge d) &=& -\rho(a\pwedge b\pwedge c\pwedge d)
   + 2 \rho(a\pwedge b) \rho(c\pwedge d)
\\&& - 2 \rho(a\pwedge c) \rho(b\pwedge d) + 2 \rho(a\pwedge d) \rho(b\pwedge c),
\end{eqnarray*}
where $a,b,c,d$ are Dirac creation or annihilation operators.
From $\rho^{-1}$ and equation (\ref{tildeu}) we can calculate
the first adapted normal products of Dirac fields.
\begin{eqnarray*}
\tilde{a} &=& a,\\
\widetilde{a\pwedge b} &=& a\pwedge b - \rho(a\pwedge b),\\
\widetilde{a\pwedge b \pwedge c} &=& a\pwedge b \pwedge c- 
     \rho(a\pwedge b) c
     +\rho(a\pwedge c) b
     -\rho(b\pwedge c) a,\\
\widetilde{a\pwedge b \pwedge c \pwedge d} &=& a\pwedge b \pwedge c \pwedge d
   - \rho(a\pwedge b) c \pwedge d
   + \rho(a\pwedge c) b \pwedge d
   - \rho(b\pwedge c) a \pwedge d
\\&&
   - \rho(a\pwedge d) b \pwedge c
   + \rho(b\pwedge d) a \pwedge c
   - \rho(c\pwedge d) a \pwedge b
   -\rho(a\pwedge b\pwedge c\pwedge d)
\\&&
   + 2 \rho(a\pwedge b) \rho(c\pwedge d)
- 2 \rho(a\pwedge c) \rho(b\pwedge d) + 2 \rho(a\pwedge d) \rho(b\pwedge c).
\end{eqnarray*}
It can be checked
that these $\tildu$ coincide with the adapted
normal products defined in reference \cite{Kutzelnigg}.

The second question posed by Kutzelnigg and Mukherjee is:
once we have defined adapted normal products, how can we express
their operator products? Again, the Hopf algebra methods yield
a complete answer. Starting from $\tildu$ and $\tildv$
we look for a 2-cocycle $\chi$ such that
\begin{eqnarray}
\tildu \tildv &=& \sum (-1)^{|u\i2||v\i1|}
\chi(u\i1,v\i1) \widetilde{u\i2\pwedge v\i2}.
\label{rtproduitdef}
\end{eqnarray}
The answer is now expected by the reader:
$\chi=(\partial \rho^{-1})\star (.|.)_+$.

At this stage, it will be useful to give a few
examples to show the kind of expressions that are
obtained by a direct calculation.
According to Lemma~\ref{lem:cobsym},
$\ZZ(u,v) = \signe{|u|\,|v|} \ZZ(v,u)$. The value
of $\ZZ(u,v)$ for the simplest elements of $\ANcal$ is,
again for Dirac fields,
\begin{eqnarray*}
\ZZ(a,b) &=& \rho(a\pwedge b),\\
\ZZ(a\pwedge b,c\pwedge d) &=& \rho(a\pwedge b\pwedge c\pwedge d)-
\rho(a\pwedge b) \rho(c\pwedge d),\\
\ZZ(a, b\pwedge c \pwedge d) &=& \rho(a\pwedge b\pwedge c\pwedge d)-
\rho(a\pwedge b) \rho(c\pwedge d)\\
  && +\rho(a\pwedge c) \rho(b\pwedge d)-\rho(b\pwedge c) \rho(a\pwedge d),
\\&=& \ZZ(a\pwedge b\pwedge c,d).
\end{eqnarray*}
Now we can calculate $\chi$ as
\begin{eqnarray*}
\chi(u,v) &=& \sum \ZZ(u\i1,v\i2) (u\i2 | v\i1)_+
=
 \sum \ZZ(u\i2,v\i1) (u\i1 | v\i2)_+.
\end{eqnarray*}
A few example computations might be appropriate 
\begin{eqnarray*}
\chi(a,b)  &=& \rho(a\pwedge b) +(a|b)_+,\\
\chi(a\pwedge b\pwedge c,d) &=& \ZZ(a\pwedge b\pwedge c , d) ,\\
\chi(a\pwedge b,c\pwedge d) &=& \ZZ(a\pwedge b,c\pwedge d)
  + (a\pwedge b|c\pwedge d)_+ - \rho(a\pwedge c) (b|d)_+
\\&&\hspace*{-7mm}+ \rho(a\pwedge d) (b|c)_+ + \rho(b\pwedge c) (a|d)_+ 
- \rho(b\pwedge d) (a|c)_+.
\end{eqnarray*}

Finally, we define the product of $\tildu$ and $\tildv$ by
$\tildu \tildv = \sum \ZZ({u\i2},{v\i1}) \,\widetilde{u\i1 v\i2}$.
This operator product satisfies (\ref{rtproduitdef}) and a few simple
examples of it are 
\begin{eqnarray*}
\tilde{a} \tilde{b} &=& \widetilde{a\pwedge b} + \rho(a\pwedge b) +(a|b)_+,\\
(\widetilde{a \pwedge b}) \tilde{c} &=& 
\widetilde{a\pwedge b\pwedge c} -(a|c)_+ \tilde{b} + (b|c)_+ \tilde{a}
+ \rho(b\pwedge c) \tilde{a} - \rho(a\pwedge c) \tilde{b}.
\end{eqnarray*}
As illustrated in reference \cite{Kutzelnigg}, the combinatorial
complexity of the products increases very quickly. The Hopf algebraic
concepts provide powerful tools to manipulate these products.

Moreover, this example coming from quantum chemistry shows that
the convolution of a Laplace pairing by a 2-coboundary 
is a very natural quantum object. Lemma~\ref{lem:2cocycle} says
that all 2-cocycles can be obtained by such
a convolution.  The adapted normal product can be
understood as inducing a transformation of the augmentation.
This parallels the situation discussed in Section~\ref{sec:cohomtop},
as exhibited in equation (\ref{eq:freen}).

The basis change defined by equation (\ref{tildeu})
appears not only in quantum chemistry, but also
in the context of quantum field theory in curved space-time.
In Ref.\cite{Hollands} (Theorem~5.1), it is shown
that two Wick monomials that can
define a quantum field theory in curved space-time
are related by equation (\ref{tildeu}).

\section{Conclusions and outlook}

We conclude this paper by sketching possible future developments and
applications of the presented framework.

From a computational point of view, the Hopf algebra structure
presented in this paper was already used to solve two old problems
of many-body theory: the hierarchy of Green functions for systems
with initial correlation \cite{BrouderDeg} and the many-body
generalization of the crystal-field method \cite{BrouderXfield}.
More results along these lines might be expected.

In an interesting parallel development
Kreimer discovered that the combinatorics of Bogoliubov's recursion
formula of renormalization and Zimmermann's solution can be expressed
in Hopf algebraic
terms \cite{Kreimer}. This was further elaborated by Connes, Kreimer
and Pinter \cite{Kreimer,CKI,PinterHopf}. In particular, it was also
shown how this leads to computational improvements compared
to traditional methods \cite{BrKr:pbresum}.
It should be possible to connect this Hopf algebraic structure
associated with the perturbative Feynman diagram expansion with the
algebraic framework introduced in this paper.
Indeed, Pinter's work might be seen as pointing in this direction
\cite{PinterHopf}. More concrete steps for establishing such a
connection were performed in \cite{BrouderSchmitt}.

Looking back at Section~\ref{sec:twist} there are really three
ingredients to the twist, the Hopf superalgebra $H$, the comodule
(super)algebra $A$ and the 2-cocycle $\chi$. In the case explored in
this paper, however, $A$ and $H$ are really the same, namely the
algebra of normal ordered field operators
(Section~\ref{sec:twistfield}). However, the fact that we obtain
deformations of this (super)algebra only depends on the choice of
$A$. In this respect the fact that $H$ may also be taken to be this
superalgebra (extended to a Hopf superalgebra) is an
``accident''.

One may ask whether a different choice for $H$ (and thus also for the
cocycle) may lead to any interesting constructions for QFT. This
is indeed the case as evidenced by \cite{Oe:nctwist}. There, in
essentially the same twist deformation construction for QFT the Hopf
algebra $H$ was chosen to correspond to the group of translations of
Minkowski space. A certain cocycle then yields QFT
on noncommutative space-times of the Moyal type.

Thus, the same twist procedure unifies a priori rather different
and unrelated constructions in QFT. It seems likely that there is
rather more to discover. It should be mentioned in this context that
it is straightforward to consider the Hopf algebraic
analogues of group products such as direct or semidirect
ones. Moreover, in
general the twist does not leave $H$ invariant. In particular, it can
lead to genuine quantum group symmetries even if the initial objects
are group symmetries. See \cite[Section~4.4]{Oe:nctwist} for examples.

Another interesting direction for generalizing our approach is to
non-linear QFT. To this end notice that $H$ (and $A_N$) may be seen as
the algebra
of polynomial functions on field configurations. Considering a
non-linear field theory the analogue would be an algebra of functions
on field configurations generated by suitably chosen modes. At least for
compact group target spaces this is rather straightforward by
employing the Peter-Weyl decomposition. Of course $H$ would no longer
be cocommutative if the respective group is non-abelian. This implies
for example that part of the cohomology theory no longer
works. However, the crucial part, namely the Drinfeld twist
generalizes to this case. This offers perspectives to describe
quantized non-linear QFTs through our approach.

Not elaborated in the course of this work was the intimate 
connection of Hopf algebraic methods to combinatorics and group
representations. The invariant theoretic content of Hopf superalgebraic 
methods was studied in \cite{Grosshans}. The connection of Hopf algebra
cohomology to branching laws, group representations and symmetric 
functions was investigated in \cite{FauserJarvis}. From this contacts 
to other fields of mathematics one expects further insight into the 
structure of QFT. This will be considered elsewhere.

 
\subsection*{Acknowledgments}

Ch.~B.\ is very grateful to Romeo Brunetti and 
Klaus Fredenhagen for an enlightening discussion.
R.~O.\ is grateful to Allen Hirshfeld for the invitation to give a
seminar on this material at Dortmund University in January 2003 and to
Florian Scheck for the invitation to give a Hauptvortrag on this subject at
the annual DPG conference in Hannover in March 2003.
He was supported through a Marie Curie Fellowship from the European
Union. B.~F.\ would like to thank Peter D. Jarvis for fruitful discussions
and the possibility to present the material in a series of lectures at 
the University of Hobart Tasmania, May--July 2003.
We thank Iana Anguelova for bringing 
Borcherds's paper to our attention.
 
\appendix

\section{Introduction to Hopf $*$-superalgebras}
\label{sec:superalg}  

In this appendix we give a detailed list of the definitions 
and main properties of Hopf superalgebras because this information 
is scattered in the literature, using various incompatible conventions 
and notations. Standard references for Hopf algebras and associated 
structures are \cite{MilnorMoore,Swe:hopfalg,Majid,Kassel}, and the super
case is emphasized in
\cite{Pittnerbook}. 
All vector spaces are over the field $\mathbb{C}$ of
complex numbers.

\subsection{Hopf superalgebras}
\label{sec:Hopfsuperalgebras}
A vector space $H$ is a \emph{super vector space} if it can be written
as $H=H_0 \oplus H_1$. If an element $a$ is in either
$H_0$ or $H_1$, we say that it is homogeneous.
If $a\in H_0$ (resp. $a\in H_1$), we say that it is even 
(resp. odd) and its parity is $|a|=0$ (resp. $|a|=1$).
The concept of super vector space enables us to consider fermions and
bosons on the same footing.
A super vector space $H$ is a \emph{superalgebra} if it is endowed with an
associative product and a unit $1 \in H_0$ and if
$|a b|=|a|+|b|$ modulo 2 for any homogeneous elements $a$ and $b$ in $H$.
A superalgebra $H$ is a \emph{Hopf superalgebra} if it is endowed with 
a coproduct $\cop: H \longrightarrow H \tens H$, a counit 
$\cou: H \longrightarrow \mathbb{C}$ and an antipode 
$\antip: H \longrightarrow H$, such that: 
\begin{itemize}
\item $\cop$ is a graded coproduct, i.e. for any homogeneous element $a$ 
   of $H$, in $\Delta a = \sum a\i1 \tens a\i2$ (using Sweedler's
   notation) all $a\i1$ and $a\i2$ are homogeneous and
   $|a|=|a\i1|+|a\i2|$.
\item $\cop$ and $\cou$ are graded algebra morphisms and $\antip$ is a graded 
algebra anti-morphism, i.e.
\begin{eqnarray*}
\cop(a b) &=& \sum (-1)^{|a\i2| |b\i1|} a\i1 b\i1\tens a\i2 b\i2, \\ 
\cou(ab) &=& \cou(a)\cou(b), \\ 
\antip(ab) &=& \signe{|a||b|} \antip(b)\antip(a);
\end{eqnarray*}
\item $\cop$, $\cou$ and $\antip$ are unital maps, i.e. 
\begin{eqnarray*}
\cop(1) = 1 \tens 1, \quad 
\cou(1) = 1 , \quad
\antip(1) = 1; &&  
\end{eqnarray*}
\item $\cop$ is coassociative, i.e. 
\begin{eqnarray*}
(\cop\tens\id)\circ\cop (a) &=& (\id\tens\cop)\circ\cop (a) \quad
\equiv \Delta^2(a) = \sum a\i1\tens a\i2\tens a\i3;   
\end{eqnarray*}
\item $\cou$ is counital, i.e. 
\begin{eqnarray*}
\sum \cou(a\i1) a\i2 = \sum a\i1 \cou(a\i2) &=& a ; 
\end{eqnarray*}
\item $\antip$ is counital, i.e. $\cou(\antip(a)) = \cou(a)$, 
and satisfies the identity 
\begin{eqnarray*}
\sum \antip(a\i1) a\i2 = \sum a\i1 \antip(a\i2) &=& \cou(a)1 .  
\end{eqnarray*}
\end{itemize}
Under these assumptions, the antipode is also a graded coalgebra 
anti-morphism, that is 
\begin{eqnarray*}
\cop \antip(a) = \sum \antip(a)\i1\tens \antip(a)\i2 &=& 
\sum \signe{|a\i1|\,|a\i2|} \antip(a\i2)\tens \antip(a\i1).
\end{eqnarray*}

A Hopf superalgebra $H$ is \emph{graded commutative} if 
$a b = (-1)^{|a||b|} b a$, and it is \emph{graded cocommutative} if 
\begin{eqnarray*}
\cop a = \sum a\i1 \tens a\i2
&=& \sum \signe{|a\i1|\,|a\i2|} a\i2 \tens a\i1,
\end{eqnarray*}
for any homogeneous $a,b \in H$. If a Hopf superalgebra $H$ is graded
commutative
or graded cocommutative, then $\antip (\antip (a))=a$ for any $a \in H$. 

A Hopf superalgebra $H$ is graded if there
are super vector spaces $H_n$ for $n\in\open{N}$ such that
\begin{eqnarray*}
H &=& \bigoplus_{n\in \open{N}} H_n.
\end{eqnarray*}
If $a\in H_n$ we say that $a$ is a homogeneous element of \emph{degree}
$n$,
and we denote the degree of $a$ by $\grad(a)$. Moreover, the degree is an
algebra map, i.e. if $a\in H_n$ and $b\in H_m$ then $a b\in
H_{n+m}$, and a coalgebra map, i.e. if $a\in H_n$ and $\cop a=\sum a\i1
\tens a\i2$ then $a\i1$ and $a\i2$ are homogeneous elements and
$\grad(a\i1)+\grad(a\i2)=n$. Finally, if $a$ is a homogeneous element,
then $\cou(a)=0$ if $\grad(a)>0$ and $\cou(1)=1$.


\subsection{Hopf $*$-superalgebra}
\label{sec:starsuper}
In quantum field theory, the adjoint operator
$a\mapsto a^*$ plays a prominent role.
Its existence is one of the basic principles of axiomatic quantum (field)
theories \cite{Haag,Bratteli2,Steinmann}. It is tied to the 
definition of a
positive quantum field state (i.e. a positive continuous linear
functional such that
$\rho(1)=1$ and $\rho(u u^*)\ge 0$). Such a $\rho$ allows via the 
GNS construction the reconstruction of a Hilbert space picture.

Therefore, it is important to specify the interplay between
the adjoint operator and the Hopf superalgebra structure.
This is done abstractly be defining a star-operation
as a bijection $*$: $H\rightarrow H$ such that
$(a^*)^*=a$ for any $a\in H$,
$(\lambda a+ \mu b)^*=\bar\lambda a^* + \bar\mu b^*$ for any
$a$, $b$ in $H$ and any complex numbers $\lambda$ and
$\mu$ (with complex conjugate $\bar\lambda$ and $\bar\mu$).
We use De Witt's convention\footnote{In the mathematical literature,
one finds also
$(ab)^*= \signe{|a|\,|b|}b^* a^*$
(e.g. \cite{Cartier,Pittner}). }:
$(ab)^*= b^* a^*$ because it is compatible with
the interpretation of $*$ as the Hermitian adjoint
of an operator.
The action of the star operation on the tensor product is
$(a\otimes b)^* = \signe{|a|\,|b|} a^* \otimes b^*$.
The compatibility of the star operation with the parity
is $|a^*|=|a|$.
The compatibility of the star operation with the coproduct is
\begin{eqnarray*}
\Delta (a^*) &=& \sum a^*\i1\otimes a^*\i2
= (\Delta a)^*
= \sum \signe{|a\i1|\,|a\i2|} {a\i1}^*\otimes {a\i2}^*.
\end{eqnarray*}
The compatibility of the star operation with the counit is
$\counit(a^*)=(\counit(a))^*$.
The compatibility
with the antipode is
\begin{eqnarray*}
\antip(\antip(a^*)^*)= a.
\end{eqnarray*}
Finally, when the Hopf superalgebra is graded,
the star operation must be compatible with this grading:
$\grad(a^*)=\grad(a)$.

\subsection{The symmetric Hopf superalgebra}
\label{sec:Example}

If $V$ is a vector space, the symmetric Hopf algebra $S(V)$ was
described in pedagogical detail in our paper \cite{BrouderOecklI}.
Here we consider the symmetric Hopf superalgebra $\Sym(V)$ where
$V$ is a super vector space (see \cite[Appendix~2]{Eisenbud}).
Let $V=V_0 \oplus V_1$ be a super vector space over $\mathbb{C}$, and denote 
by $|v|$ the parity of a homogeneous element $v \in V$. 
Let $T(V) = \oplus_{n=0}^\infty V^{\otimes n}$ be the tensor algebra on $V$, 
with the tensor (free) product $\otimes$, and unit 
$1 \in \mathbb{C} = V^{\otimes 0}$. 

The \emph{symmetric superalgebra on $V$} is the quotient of $T(V)$ by the 
supersymmetric (or graded commutative) relation, that is
\begin{eqnarray*}
\Sym(V) &=& T(V)/\langle u \otimes v - (-1)^{|u| |v|} v \otimes u \rangle, 
\end{eqnarray*}
where the elements $u,v$ are homogeneous in $V$. 
Since the ideal is generated by a homogeneous relation, the quotient $\Sym(V)$ 
is still a graded vector space, that is 
$\Sym(V) = \oplus_{n=0}^\infty \Sym^n(V)$, 
and it has homogeneous components 
$\Sym^n(V) = \sum_{p+q=n} S^p(V_0) \otimes \Lambda^q(V_1)$, where 
$S^p(V_0)$ denotes the symmetric $p$-powers on $V_0$ and $\Lambda^q(V_1)$ 
denotes the exterior (skew-symmetric) $q$-powers on $V_1$. 
Then, in the quotient $\Sym(V)$ we denote by $\pwedge$ the
concatenation product
induced by $\otimes$. Explicitly, for $u_1\pwedge\cdots\pwedge u_r \in
\Sym^r(V)$
and $v_1\pwedge\cdots\pwedge v_s \in \Sym^s(V)$ we have 
\begin{eqnarray*}
(u_1\pwedge\cdots\pwedge u_r) \pwedge (v_1\pwedge\cdots\pwedge v_s) &=& 
u_1\pwedge\cdots\pwedge u_r \pwedge v_1\pwedge\cdots\pwedge v_s \in
\Sym^{r+s}(V).
\end{eqnarray*}
The product $\pwedge$ is associative (as well as $\otimes$), unital (with unit 
$1 \in \mathbb{C} = \Sym^0$) and graded commutative, that is 
$u \pwedge v = (-1)^{|u| |v|} v \pwedge u$ for homogeneous $u,v \in V$. 

As a vector space, $\Sym(V)$ is isomorphic to $S(V_0) \otimes \Lambda(V_1)$, 
but not as an algebra. In fact, in $S(V_0) \otimes \Lambda(V_1)$ 
the product is among each tensor component (bosons with bosons, fermions 
with fermions), while in $\Sym(V)$ we may wish to multiply the two components 
(bosons with fermions). As an algebra, $\Sym(V)$ is isomorphic to 
the superalgebra $H_0\oplus H_1$,
where $H_0=S(V_0) \otimes \oplus_{n=0}^\infty \Lambda^{2n}(V_1)$ has even 
parity and $H_1=S(V_0) \otimes \oplus_{n=0}^\infty \Lambda^{2n+1}(V_1)$ 
has odd parity. For this, it suffices to check that $H_0 \pwedge H_0$ and 
$H_1 \pwedge H_1$ are subsets of $H_0$, and that $H_0 \pwedge H_1$ and 
$H_1 \pwedge H_0$ are subsets of $H_1$. 

The symmetric superalgebra $\Sym(V)$ can be endowed with a coassociative 
coproduct $\Delta: \Sym(V) \rightarrow \Sym(V) \otimes \Sym(V)$, defined 
on the generators $v \in V$ as $\Delta v = 1 \otimes v + v \otimes 1$ and 
extended to products $v_1 \pwedge\cdots\pwedge v_n$ as an algebra morphism. 
Due to the graded commutativity of the product $\pwedge$, the formula for a 
generic element of length $n$ can be explicitly given in terms of the 
$(p,n-p)$-shuffles, which are the permutations $\sigma$ on $n$ elements 
such that $\sigma(1)<\cdots<\sigma(p)$ and $\sigma(p+1)<\cdots<\sigma(n)$:
if $u=v_1\pwedge\dots\pwedge v_n$,
\begin{eqnarray}
\Delta u &=& u \otimes 1 + 1 \otimes u
\nonumber\\
&& + \sum_{p=1}^{n-1} (-1)^F v_{\sigma(1)}\pwedge \dots\pwedge
v_{\sigma(p)} \otimes
v_{\sigma(p+1)}\pwedge \dots\pwedge v_{\sigma(n)}, 
\label{Deltav1vn}
\end{eqnarray}
where
\begin{eqnarray*}
F&=& \sum_{i=1}^p \sum_{j=p+1}^n
\theta\big(\sigma(i)-\sigma(j)\big) |v_{\sigma(i)}||v_{\sigma(j)}|,
\end{eqnarray*}
with $\theta\big(\sigma(i)-\sigma(j)\big)=1$ if
$\sigma(i)>\sigma(j)$ and
$\theta\big(\sigma(i)-\sigma(j)\big)=0$ if
$\sigma(i)<\sigma(j)$.
When all operators are odd ($|v_i|=1$ for all $i$),
$(-1)^F$ is the signature $(-1)^\sigma$
of the permutation $\sigma$,
when all operators are even, $(-1)^F=1$.

The simplest example of equation (\ref{Deltav1vn}) is
\begin{eqnarray*}
\Delta(v_1 \pwedge v_2) &=& v_1 \pwedge v_2 \otimes 1 + 1 \otimes v_1
\pwedge v_2 
+ v_1 \otimes v_2 + (-1)^{|v_1||v_2|} v_2 \otimes v_1.
\end{eqnarray*}
The linear map $\cou: \Sym(V) \rightarrow \mathbb{C}$ which is the 
identity on the scalars $\mathbb{C} = V^0 \subset \Sym(V)$, and zero 
on higher degrees, is a counit for this coproduct. 
Moreover an antipode is then automatically defined by induction on the length 
of the elements. In conclusion, the symmetric superalgebra $\Sym(V)$ has 
the structure of a graded commutative Hopf superalgebra, as defined in 
Appendix~\ref{sec:Hopfsuperalgebras}.

\section{Cohomology computations for $\Sym(V)$}
\label{cohomsect}

\subsection{Cohomology groups of bosonic $\Sym(V)$}
\label{sec:cohomgrp}

In the bosonic case, i.e.\ if $V$ is purely even, we can work out the
cohomology groups of $\Sym(V)$ as follows.

The symmetric algebra $\Sym(V)$ can be seen as the universal
enveloping algebra $\mathsf{U}(V)$ of the abelian Lie algebra $V$
(with all brackets set to zero). Sweedler proves in \cite[Thm.~4.1,
page~212]{Sweedler} that the Hopf algebra cohomology
$H^\bullet(\Sym(V))$ is isomorphic to the Hochschild cohomology
$HH^\bullet(\mathsf{U}(V))$. This is known to be isomorphic to the
Chevalley-Eilenberg cohomology $H^\bullet(V)$ of the Lie
algebra. Since $V$ is abelian, all coboundary operators are zero and
$H^\bullet(V)$ is easily computed:
$H^\bullet(V)=[\Lambda^\bullet(V)]^*$.
Hence $H^1(\Sym(V))=V^*$ and
$H^2(\Sym(V))=(\Lambda^2(V))^*=\Hom_{\C}(\Lambda^2(V),\C)$.

\subsection{Proofs of cohomological statements}
\label{sec:proofs}

\subsubsection*{Proof of Proposition~\ref{prop:twistcohom}}

We first show that $T:A_\eta \to A_\chi$ given by $T(a)=\sum
\rho(a\i1) a\iu2$ is a comodule map. Using graded cocommutativity we
find as required,
\begin{equation*}
\begin{split}
\sum T(a)\i1\tens T(a)\iu2 = & \sum \rho(a\i1) a\i2 \tens a\iu3\\
 = & \sum a\i1 \tens \rho(a\i2) a\iu3\\
 = & \sum a\i1 \tens T(a\iu2) .
\end{split}
\end{equation*}

We prove secondly that $T$
is a superalgebra isomorphism. We
denote the twisted product induced by $\eta$ by $\circ_\eta$ and the
twisted product induced by $\chi$ by $\circ_\chi$. Using graded
cocommutativity we find as required,
\begin{equation*}
\begin{split}
T(a\circ_\eta b) = & \sum (-1)^{|b\i1| |a\iu2|}
  \eta(a\i1,b\i1) T(a\iu2 b\iu2)\\
 = & \sum (-1)^{|a\i1| |a\i2|+|a\i1| |a\i3|+|a\i1| |a\iu4|+|a\i2| |a\i3|
  +|a\i2| |a\iu4| + |a\i3| |a\iu4|}\\
 & \partial\rho(a\i1, b\i1) \chi(a\i2, b\i2) \rho(a\i3 b\i3)
  a\iu4 b\iu4\\
 = & \sum (-1)^{|b\i2| |a\iu3|}
  \rho(a\i1) \rho(b\i1) \chi(a\i2, b\i2) a\iu3 b\iu3\\
 = & \sum (-1)^{|b\i1| |a\iu2|}
  \chi(a\i1, b\i1) T(a\iu2) T(b\iu2)\\
 = & \sum (-1)^{|T(b)\i1| |T(a)\iu2|}
 \chi(T(a)\i1, T(b)\i1) T(a)\iu2 T(b)\iu2)\\
 = & T(a)\circ_\chi T(b) .
\end{split}
\end{equation*}

For the converse direction we have $A=H$ and the coaction is the
coproduct of $H$. By assumption we have a comodule superalgebra
isomorphism $T:A_\eta \to A_\chi$. Let $\rho:H\to\C$ be defined by
$\rho\defeq \cou\circ T$. Then $\rho$ is unital,
$\rho(1)=1$. Furthermore, it is invertible with inverse given by
$\rho^{-1}=\cou\circ T^{-1}$, and thus it is a 1-cochain on $H$.
Since $T$ is a comodule map we can show using graded cocommutativity
\begin{equation*}
\begin{split}
 \sum \rho(a\i1) a\i2= & \sum \cou(T(a\i1)) a\i2 = \sum a\i1 \cou(T(a\i2))\\
 = & \sum T(a)\i1 \cou(T(a)\i2)=T(a).
\end{split}
\end{equation*}
That is, $T$ is determined by $\rho$ as in the proposition. We show
that $\eta$ and $\chi$ are cohomologous through $\rho$, i.e.\
$\eta=\partial\rho\star\chi$:
\begin{equation*}
\begin{split}
\eta(a,b) = & \sum (-1)^{|a\i1| |a\i2|+ |a\i1| |a\i3| + |a\i2| |a\i3|}\\
 & \eta(a\i1,b\i1) \rho(a\i2 b\i2) \rho^{-1}(a\i3 b\i3)\\
 = & \sum (-1)^{|a\i1| |a\i2|+ |a\i1| |a\i3| + |a\i2| |a\i3|}\\
 & \cou\circ T(\eta(a\i1,b\i1) a\i2 b\i2)\rho^{-1}(a\i3 b\i3)\\
 = & \sum (-1)^{|b\i1| |a\i2|}
 \cou\circ T(a\i1\circ_\eta b\i1) \rho^{-1}(a\i2 b\i2)\\
 = & \sum (-1)^{|b\i1| |a\i2|}
 \cou(T(a\i1)\circ_\chi T(b\i1)) \rho^{-1}(a\i2 b\i2)\\
 = & \sum (-1)^{|b\i2| |a\i3|}
 \rho(a\i1) \rho(b\i1) \cou(a\i2\circ_\chi b\i2)
 \rho^{-1}(a\i3 b\i3)\\
 = & \sum (-1)^{|b\i1| |a\i2|} \partial\rho(a\i1,b\i1) \chi(a\i2,b\i2) .
\end{split}
\end{equation*}
This completes the proof.

\subsubsection*{Proof of Lemma~\ref{lem:1coch}}

We prove that the map $\mu:Z^1\times N^1\to C^1$ given by the convolution
product is bijective.

Let $\rho\in C^1$. Define $\zeta:\Sym(V)\to\C$ as
follows. Set $\zeta(1)=1$, set $\zeta(v)=\rho(v)$ for $v\in V$ and
extend $\zeta$ to all of $\Sym(V)$ as an algebra homomorphism $\zeta(a\pwedge
b)=\zeta(a)\zeta(b)$. (Note that $\zeta$ is automatically graded since
$\rho$ is graded.)
$\zeta$ has a convolution inverse
$\zeta^{-1}(a)= \zeta(\gamma(a))$ and is thus a 1-cocycle. Now define
the 1-cochain $\eta\defeq \zeta^{-1} \star \rho$. Since
$\eta(v)=\zeta^{-1}(1)\rho(v) +
\zeta^{-1}(v)\rho(1)=\rho(v)-\rho(v)=0$ for all $v\in V$, $\eta$ is in
$N^1$. By construction, $\rho=\zeta\star\eta$ and thus $\mu$ is
surjective.

Now take $\tilde{\zeta}\in Z^1$ and $\tilde{\eta}\in N^1$. Define
$\rho\defeq \tilde{\zeta}\star\tilde{\eta}$ and construct as above
$\zeta, \eta$ from $\rho$. Since
$\rho(v)=\tilde{\zeta}(1)\tilde{\eta}(v)
+\tilde{\zeta}(v)\tilde{\eta}(1)=\tilde{\zeta}(v)$ for $v\in V$ we have by
construction of $\zeta$ that $\zeta(v)=\tilde{\zeta}(v)$ for $v\in
V$. As both $\zeta$ and $\tilde{\zeta}$ are algebra homomorphisms they
must coincide. Consequently $\eta=\zeta^{-1} \star
\rho=\tilde{\zeta}^{-1}\star
\tilde{\zeta}\star\tilde{\eta}=\tilde{\eta}$. 
This shows that $\mu$ is injective.


\subsubsection*{Proof of Lemma~\ref{lem:cobsym}}

We show that a 2-cocycle $\chi$ on $\Sym(V)$ is a 2-coboundary 
if and only if it is symmetric, i.e. $\chi(a,b)=(-1)^{|a|
|b|}\chi(b,a)$ for all $a,b$ in $\Sym(V)$. 

Let $\chi$ be a 2-coboundary, then there is a 1-cochain $\rho$ such that
\begin{equation*}
\chi(a,b) = \partial \rho(a,b) = 
\sum \rho(a\i1) \rho(b\i1) \rho^{-1}(a\i2\pwedge b\i2). 
\end{equation*}
Since the symmetric product $\pwedge$ is graded commutative, the
expression on the right is symmetric and $\chi(a,b) = (-1)^{|a|
  |b|}\chi(b,a)$.
 
For the reciprocal statement, suppose that $\chi$ is a symmetric
2-cocycle. We define a 1-cochain on $S(V)$ by induction on $\grad(v)$.
Set $\rho(1)=1$, $\rho(v)=0$ for all $v\in V$, and $\rho(u \pwedge v) =
\chi^{-1}(u,v)$ for all $u,v\in V$
(which is well defined because $\chi$ is symmetric). 
Now assume that $\rho$ is defined on all elements up to degree $\leq
n$. For $a, b\in \Sym(V)$ with $\grad(a)+\grad(b)= n+1$ and 
$\grad(a), \grad(b)\ge 1$ (for $a$ or $b$ of $\grad(0)$ what is to be
shown holds automatically), set 
\begin{eqnarray}
\label{defrho}
\rho(a \pwedge b) &=& \sum \chi^{-1}(a\i1,b\i1) \rho(a\i2)\rho(b\i2).
\end{eqnarray}  
Then $\rho$ is well defined because $\chi^{-1}$ is symmetric, 
and therefore $\rho(a \pwedge b) = \rho((-1)^{|a| |b|} b \pwedge a)$, and 
because 
$\rho(a \pwedge (b\pwedge c)) = \rho((a \pwedge b) \pwedge c)$ if 
$\grad(a)+\grad(b)+\grad(c)= n+1$ 
and $\grad(a), \grad(b), \grad(c)\ge 1$. 
To show the latter equality, we use equation (\ref{defrho}), the
2-cocycle condition (\ref{eq:cocycleid}) and coassociativity of the
coproduct to write
\begin{eqnarray*}
\rho(a\pwedge (b \pwedge c)) 
&=& \sum (-1)^{|c\i1| |b\i2|}\chi^{-1}(a\i1,b\i1\pwedge c\i1)
 \rho(a\i2) \rho(b\i2\pwedge c\i2) \\ 
&=& \sum (-1)^{|c\i1| |b\i4|+ |c\i2| |b\i4| + |b\i1| |b\i2| + |c\i1|
   |c\i2|}
 \chi^{-1}(a\i1 \pwedge b\i1, c\i1) \\
&& \hspace{1cm} \chi^{-1}(a\i2,b\i2)
 \chi(b\i3,c\i2)  \rho(a\i3) \rho(b\i4 \pwedge c\i3) \\ 
&=& \sum (-1)^{|b\i1| |b\i2|} \chi^{-1}(a\i1 \pwedge b\i1, c\i1)
 \chi^{-1}(a\i2,b\i2) \\ 
&& \hspace{1cm} \rho(a\i3) \rho(b\i3) \rho(c\i2) \\ 
&=& \sum (-1)^{|b\i1| |b\i2|}\chi^{-1}(a\i1 \pwedge b\i1, c\i1)
 \rho(a\i2 \pwedge b\i2) \rho(c\i2) \\ 
&=& \rho((a \pwedge b) \pwedge c). 
\end{eqnarray*}
Finally, inverting (\ref{defrho}) to obtain 
$\rho^{-1}(a \pwedge b) = \sum \rho^{-1}(a\i1)\rho^{-1}(b\i1)
\chi(a\i2,b\i2)$, 
we can easily show that $\partial \rho = \chi$. In fact,
\begin{eqnarray*}
\partial \rho(a, b) &=& \sum \rho(a\i1) \rho(b\i1)
\rho^{-1}(a\i2\pwedge b\i2) \\ 
&=& \sum \rho(a\i1) \rho(b\i1) \rho^{-1}(a\i2)\rho^{-1}(b\i2)
\chi(a\i3,b\i3) \\ 
&=& \sum \counit(a\i1) \counit(b\i1) \chi(a\i2,b\i2) = \chi(a,b). 
\end{eqnarray*}


\subsubsection*{Proof of Lemma~\ref{lem:2cocycle}}

We first consider an auxiliary Lemma.
\begin{lem}
\label{lem:cocsym}
Let $\chi$ be a 2-cocycle. If $\chi$ is symmetric on $V\tens V$, then
$\chi$ is symmetric on $\Sym(V)\tens\Sym(V)$, i.e.\ $\chi\in
Z^2_\mathrm{sym}$.
\end{lem}
\begin{proof}
Let $\chi$ be a 2-cocycle such that $\chi(u,v)=(-1)^{|u|
  |v|}\chi(v,u)$.
We prove by induction that $\chi\in Z^2_\mathrm{sym}$.
Suppose that $\chi$ is symmetric on all elements $a \tens b$ with 
$\grad(a)+\grad(b) \leq n$, and let $\grad(a)+\grad(b)+\grad(c) = n+1$, 
with $\grad(a),\grad(b),\grad(c)\geq 1$.
Using the graded commutativity of the symmetric product $\pwedge$, the
2-cocycle condition (\ref{eq:cocycleid}) and the
induction hypothesis, we obtain: 
\begin{eqnarray*}
\chi(a,b \pwedge c) &=&  \sum (-1)^{|a\i1| |a\i2|+|b\i1| |b\i2| + |b\i1|
  |b\i3|}\\ 
&& \chi^{-1}(b\i1,c\i1)\chi(a\i1,b\i2)\chi(a\i2\pwedge b\i3, c\i2)\\
&=&  \sum (-1)^{|a\i1| |a\i2|+|b\i1| |b\i2| + |b\i1|
  |b\i3| + |a\i1| |b\i2| + |a\i2| |b\i3|}\\ 
&& \chi^{-1}(b\i1,c\i1)\chi(b\i2,a\i1)\chi(b\i3\pwedge a\i2, c\i2)\\
&=&  \sum (-1)^{|a\i2| |b\i2|+|b\i1| |b\i2| + |c\i2|
  |c\i3|}\\ 
&& \chi^{-1}(b\i1,c\i1)\chi(a\i1,c\i2)\chi(b\i2, a\i2\pwedge c\i3)\\
&=&  \sum (-1)^{|a\i1| |c\i2| +|a\i2| |b\i2|+|a\i2| |c\i3|+|b\i1|
  |b\i2| + |c\i2| |c\i3|}\\ 
&& \chi^{-1}(b\i1,c\i1)\chi(c\i2,a\i1)\chi(b\i2, c\i3 \pwedge a\i2)\\
&=& (-1)^{|a| |b \pwedge c|}\chi(b \pwedge c,a).  
\end{eqnarray*}
This completes the proof.
\end{proof}

We are now ready to prove the main statement that the map
$\mu:B^2\times R^2_\mathrm{asym}\to Z^2$ given
by the convolution product is bijective.

Let $\chi$ be a 2-cocycle. Define the Laplace pairing $\lambda$ by
$\lambda(u,v)\defeq \frac{1}{2}(\chi(u,v)-(-1)^{|u| |v|}\chi(v,u))$
for all $u,v$ in $V$ extended to $\Sym(V)$ by (\ref{eq:pair1}) and
(\ref{eq:pair2}). $\lambda$ is an antisymmetric Laplace pairing
according to the definition (\ref{eq:lapasym}).
By Lemma~\ref{lem:lapcoc} $\lambda$ is a 2-cocycle
and hence $\sigma\defeq \chi\star\lambda^{-1}$ is also a
2-cocycle. Note that the inverse of $\lambda$ is the Laplace pairing
defined by $\lambda^{-1}(u,v)=-\lambda(u,v)$. Thus, $\sigma$ evaluated
on $V\tens V$ yields $\sigma(u,v)=\sum (-1)^{|v\i1| |u\i2|}
\chi(u\i1, v\i1) \lambda^{-1}(u\i2,
v\i2)=\chi(u,v)-\lambda(u,v)=\frac{1}{2}(\chi(u,v)+(-1)^{|u|
  |v|}\chi(v,u))$. That is, $\sigma$ is symmetric on $V\tens V$. By
Lemma~\ref{lem:cocsym} this implies that $\sigma$ is symmetric on all
of $\Sym(V)$ and thus by Lemma~\ref{lem:cobsym} a 2-coboundary. By
construction $\chi=\sigma\star \lambda$, i.e.\ $\chi$ can be written
as a product of a 2-coboundary $\sigma$ and an antisymmetric Laplace
pairing $\lambda$. Hence $\mu$ is surjective.

Now take $\tilde{\sigma}\in B^2$ and $\tilde{\lambda}\in
R^2_\mathrm{asym}$. Define the 2-cocycle $\chi\defeq \tilde{\sigma}\star
\tilde{\lambda}$. Construct $\sigma\in B^2$ and
 $\lambda\in R^2_\mathrm{asym}$ out of
$\chi$ as
above. Then for $u,v$ in $V$, $\chi(u,v)=\sum (-1)^{|v\i1| |u\i2|}
\tilde{\sigma}(u\i1,v\i1)
\tilde{\lambda}(u\i2,v\i2)=\tilde{\sigma}(u,v) + \tilde{\lambda}(u,v)$.
Hence
$\lambda(u,v)=\frac{1}{2}(\chi(u,v)-(-1)^{|u|
  |v|}\chi(v,u))
=\frac{1}{2}(\tilde{\sigma}(u,v)+\tilde{\lambda}(u,v)
-(-1)^{|u| |v|}(\tilde{\sigma}(v,u)+\tilde{\lambda}(v,u)))
=\tilde{\lambda}(u,v)$. We have used that $\tilde{\lambda}$ is
antisymmetric
by assumption while $\tilde{\sigma}$ is symmetric by
Lemma~\ref{lem:cobsym}. Since $\lambda$ and $\tilde{\lambda}$ are
Laplace pairings coinciding on $V\tens V$ they must be identical.
Consequently $\sigma=\chi\star\tilde{\lambda}^{-1}=\tilde{\sigma}\star
\tilde{\lambda}\star \tilde{\lambda}^{-1}=\tilde{\sigma}$.
This shows that $\mu$ is injective.

\bibliographystyle{amsordx}
\bibliography{qed}
\end{document}